\documentclass[preprint,12pt]{elsarticle}
\usepackage{natbib}

\usepackage{siunitx}

\usepackage[export]{adjustbox}
\usepackage{graphicx}

\usepackage{makecell} % For easier line breaks in the table
\usepackage{pdflscape} % Use this to rotate the page in the PDF

\usepackage{amssymb}
\usepackage{amsmath}

\journal{arXiv}
\begin{document}

\begin{frontmatter}

%% Title, authors and addresses

%% use the tnoteref command within \title for footnotes;
%% use the tnotetext command for theassociated footnote;
%% use the fnref command within \author or \address for footnotes;
%% use the fntext command for theassociated footnote;
%% use the corref command within \author for corresponding author footnotes;
%% use the cortext command for theassociated footnote;
%% use the ead command for the email address,
%% and the form \ead[url] for the home page:
%% \title{Title\tnoteref{label1}}
%% \tnotetext[label1]{}text
%% \author{Name\corref{cor1}\fnref{label2}}
%% \ead{email address}
%% \ead[url]{home page}
%% \fntext[label2]{}
%% \cortext[cor1]{}
%% \affiliation{organization={},
%%             addressline={},
%%             city={},
%%             postcode={},
%%             state={},
%%             country={}}
%% \fntext[label3]{}

\title{Numerical Investigation of Radiative Transfers Interactions with Material Ablative Response for Hypersonic Atmospheric Entry}

%% use optional labels to link authors explicitly to addresses:
\author[1]{Vincent Le Maout\corref{cor1}}
\author[2]{Sung Min Jo \corref{cor1}}
\author[3]{Alessandro Munaf\`o}
\author[3]{Marco Panesi \corref{cor2}}

\affiliation[1]{Space Vehicle Design and Hypersonics (SVDH) Lab, Department of Aerospace System Engineering, Sejong University,  Republic of Korea}
\affiliation[2]{Department of Aerospace Engineering, Korea Advanced Institute of Science and Technology, Daejeon, 34141, Republic of Korea}
\affiliation[3]{Center for Hypersonics & Entry System Studies, University of Illinois at Urbana-Champaign,IL, USA}

\cortext[cor1]{V. Le Maout and S. M. Jo equally contributed.} %% E
\cortext[cor2]{Corresponding author, mpanesi@illinois.edu} %% E

\begin{abstract} Radiative transfer interactions with material ablation are critical contributors to vehicle heating during high-altitude, high-velocity atmospheric entry. However, the inherent complexity of fully coupled multi-physics models often necessitates simplifying assumptions, which may overlook key phenomena that significantly affect heat loads, particularly radiative heating. Common approximations include neglecting the contribution of ablation products, applying simplified frozen wall boundary conditions, or treating radiative transfer in a loosely coupled manner. 
This study introduces a high-fidelity, tightly coupled multi-solver framework designed to accurately capture the multi-physics challenges of hypersonic flow around an ablative body. The proposed approach consistently accounts for the interactions between shock-heated gases, surface material response, and radiative transfer. Our results demonstrate that including radiative heating in the surface energy balance substantially influences the ablation rate. Ablation products are shown to absorb radiative heat flux in the vacuum-ultraviolet spectrum along the stagnation line, while strongly emitting in off-stagnation regions. These findings emphasize the necessity of a tightly coupled multiphysics framework to faithfully capture the complex, multidimensional interactions in hypersonic flow environments, which conventional, loosely coupled models fail to represent accurately. \end{abstract}

%%Graphical abstract
%\begin{graphicalabstract}
%%\includegraphics{grabs}
%\end{graphicalabstract}

%%%Research highlights
%\begin{highlights}
%\item A strongly coupled multi-solver framework has been proposed
%\item Surface energy balance with radiative heating critically changes material ablation rate
%\item Multi-dimensional flow-material-radiation interactions distinct from conventional stagnation line analysis demonstrated
%\end{highlights}
    
\begin{keyword}
Hypersonic Flow, Radiative Transfer, Material Ablation, Multi-physics Coupling
%% keywords here, in the form: keyword \sep keyword

%% PACS codes here, in the form: \PACS code \sep code

%% MSC codes here, in the form: \MSC code \sep code
%% or \MSC[2008] code \sep code (2000 is the default)

\end{keyword}

\end{frontmatter}

%% \linenumbers

%% main text
\section{Introduction}
\label{sec:intro}

The study of hypersonic flows during high-speed atmospheric entry is inherently multi-physics in nature, involving the interaction of various physical phenomena across multiple research domains \citep{Barbante2004, Gnoffo1999}. Hypersonic flows are characterized by converting a significant portion of kinetic energy into internal energy due to the strong shock waves around the entry vehicle, resulting from compressibility effects. This process creates a shock layer near the vehicle's surface. At high Mach numbers, changes in thermodynamic properties, such as temperature and pressure, can cause the flow to reach a thermochemical non-equilibrium state \cite{Josyula2015, Park1990}. In this state, significant excitation of the internal energy modes of flow particles occurs, leading to multiple relaxation pathways in the post-shock region. These pathways include chemical reactions such as ionization, dissociation, and recombination \cite{Josyula2015}, along with radiative transitions \cite{Johnston2008} and gas-surface interactions \cite{Marschall2015}.

Understanding and accurately quantifying heat transfer processes in the non-equilibrium hypersonic environment surrounding entry vehicles is critical to ensuring the safety of the payload during descent. To protect internal components from extreme temperatures and highly reactive aerothermal environments, \emph{Thermal Protection Systems} (TPS) are used. The design of TPS depends on the accurate characterization of the conditions encountered during atmospheric entry \cite{Uyanna2020}. TPS typically consists of carbon-fiber-reinforced phenolic composites in high-speed entries, such as super-orbital trajectories, which mitigate the high incoming heat flux through endothermic decomposition of internal components exposed to elevated temperatures. This process occurs via pyrolysis of the heat shield material and subsequent ablation of the fibers \cite{Meurisse2018}. The resulting charred gas flux is expelled into the boundary layer, reducing the incoming convective heating. However, ablation products also interact with the initial components of the hypersonic boundary layer, triggering additional chemical reactions that must be considered to accurately model gas-surface interactions \cite{Johnston2012}.

In the context of radiative transfer within the flow, introducing carbon species resulting from surface ablation has been shown to significantly affect the boundary layer near the wall \cite{Park2001}. Previous studies have identified two concurrent interaction mechanisms when considering surface ablation and radiation in atmospheric entry simulations.
Firstly, introducing strongly radiating species, such as CO and CN, originated by wall ablation products, leads to partial absorption of shock layer radiative emissions.
 This absorption reduces the amount of radiative energy reaching the vehicle surface, leading to a net reduction in radiative heating \cite{Moss1976}, but also results in a net increase in convective heating \cite{Park2001}. Numerical simulations of different entry scenarios have illustrated this effect. For example, the introduction of charred gas blowing at the surface led to a reduction in radiative heat flux of approximately 10\% for the Apollo 4 mission \cite{Park2001}, and by up to 40\% for Mars return missions \cite{Johnston2009}.

Conversely, studies involving higher Earth entry speeds have reported that ablation products can serve as a source of additional emissions, particularly in the infrared region of the spectrum \cite{Olynick1999}. The underlying mechanism for this phenomenon is attributed to enhanced radiative emissions from carbonaceous species in the hot boundary layer. As a result, the relative contribution of radiative heat flux from ablation products is closely linked to the trajectory point and the surface blowing rate, a function of entry velocity. For example, in the Stardust entry, numerical simulations predict an increase in radiative heat flux of up to 38\% when emissions from ablative products are accounted for \cite{Johnston2014}. Although to a lesser extent, a sensible increase in the radiative heat flux was also observed for the MUSE-C entry case \cite{Doihara2004}.
Thus, in conclusion, the nontrivial interaction between ablation species and radiative transfer and their impact on surface radiative heating requires further investigation, especially for higher entry speed applications, as emphasized in NASA's technology road maps for future missions \cite{NASA2015}.

One of the primary challenges in analyzing high-speed entry scenarios is the significant uncertainty associated with the large number of parameters required to model the coupled problem, including those governing spectral properties, chemical kinetics, mixture properties, and material ablation responses. For example, the wide variation in finite-rate carbon ablation models available in the literature \cite{Candler2012} leads to considerable uncertainties in radiation predictions. These predictions are highly sensitive to the selected gas-surface interaction model, as uncertainties in this model affect both the boundary layer temperature and surface mass blowing rate, which, in turn, alter the radiation interaction.

Consequently, most coupled ablative-radiative hypersonic flow studies have been conducted under the assumption of an equilibrium ablative wall 
%\cite{Doihara2004}, 
\cite{Doihara2004,padovan2024extended,chiodi2022chyps}, 
where the temperature, blowing rate, and surface composition are fixed. These assumptions neglect dynamic ablation processes and spatial variations across the surface, allowing solutions to be obtained using hypersonic flow models with prescribed blowing conditions \cite{Gnoffo2010}, simulating the effects of char decomposition based on \emph{a priori} knowledge of the elemental composition of the charred gas and surface temperature. This equilibrium wall model for ablation was later abandoned by Johnston \emph{et al.} \cite{Johnston2014} in favor of a finite-rate approach. In their work, the authors, building upon the Park-5 finite-rate carbon ablation model, investigated the sensitivity of the problem to kinetic databases. Their analysis involved reducing gas-phase chemical reaction rate coefficients by an order of magnitude \cite{Johnston2014} while increasing ablation rates for the Park-5 model. It was recognized that at high entry speeds, particularly in the carbon sublimation regime, the production of C$_3$ was the most influential factor affecting convective heating due to its substantial impact on the surface blowing rate. Additionally, the study found that radiative heating is highly sensitive to variations in gas-phase kinetic rates.

To advance the state-of-the-art in the characterization of the complex interaction mechanisms between ablative material response and radiative transfer in hypersonic shock layers, we introduce a novel multi-physics solver that achieves strong coupling across multiple domains: ablation, radiation, and thermo-chemically reacting flows \cite{Jo2023b}. Unlike previous approaches, which often treat these phenomena in a loosely coupled or sequential manner, our framework integrates them in a physically consistent way. A key advancement is the incorporation of radiative heating directly into the material energy balance, enabling more accurate predictions of wall temperature and surface blowing rates, both of which critically influence heat transfer and ablation dynamics. Additionally, the framework accounts for the finite-rate nature of gas-surface interaction processes, eliminating the equilibrium assumptions that historically constrained the fidelity of these models. This inclusion marks a significant departure from traditional methodologies, providing a more comprehensive treatment of non-equilibrium conditions in the hypersonic regime. The framework leverages a set of well-validated single-physics solvers, including \texttt{PATO} \cite{Meurisse2018} for material response, \texttt{HEGEL} \cite{Munafo2020} for hypersonic flow, and \texttt{MURP} \cite{Jo2023a} for radiative transfer, each independently verified in their respective domains. These solvers are strongly coupled through the \texttt{preCICE} library \cite{Chourdakis2022}, enabling volumetric and surface-level data exchange to ensure physical consistency and dynamic feedback between all components.

The structure of this paper is as follows: Section \ref{sec:multisolver} provides an overview of the numerical approach used to compute the interactions between hypersonic flow, radiative transfer, and material response, with a detailed explanation of the multiple coupling terms that arise in the multi-physics simulations. Section \ref{Sec:Results} presents the results. In Section \ref{CoupledAbl}, we explore the fundamental coupling behavior between thermochemical non-equilibrium flows and material response by using both equilibrium and finite-rate ablation models, analyzing the sensitivity of the coupling simulations to the parameter ranges of the ablation models. Section \ref{CoupledRes} examines the impact of radiative heat transfer coupling on material response, along with the overall characterization of total aerothermal heating. Finally, Section \ref{Sec:Conclusion} summarizes the key conclusions of this study.

\section{A Multi-Solver Approach for Coupled Interactions}
\label{sec:multisolver}
\begin{figure}[h!]
    \centering
    \includegraphics[width=0.7\textwidth]{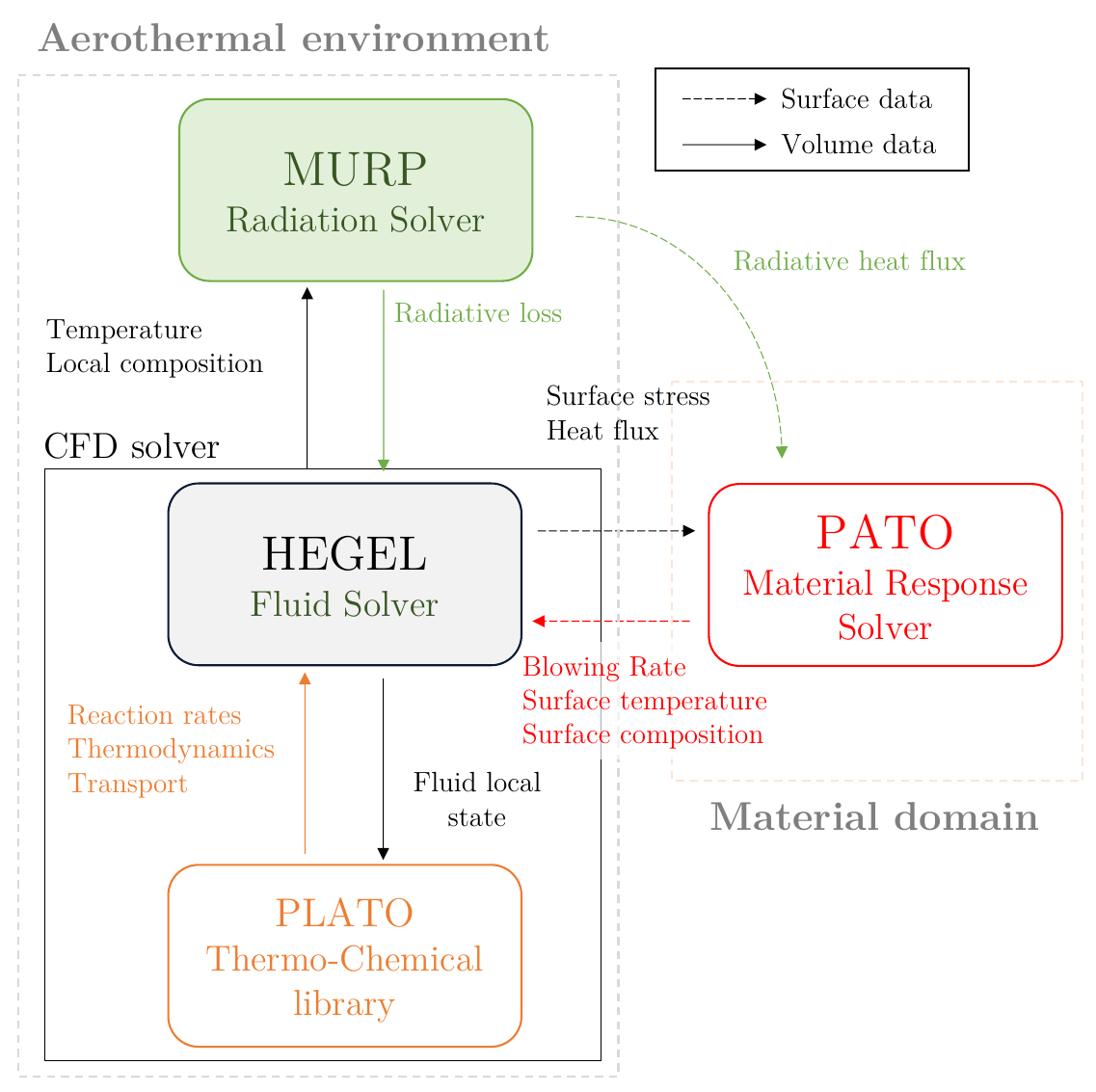}
    \caption{Multi-solver organization overview for computing radiative ablative hypersonic flow around a blunt body.}
    \label{Framework}
\end{figure}

The numerical strategy employed for efficient and accurate simulations of multi-physics coupled high-speed atmospheric entry enables the exchange of coupling terms between the dedicated solvers for each individual physics domain \cite{Jo2023b}, as illustrated in Fig. \ref{Framework}. This section describes the implementation details of the mathematical formulation, including the underlying assumptions, and emphasizes the aspects of data communication between the solvers.

% Figure-1 in MS Word

\subsection{Radiative Transfers}
The radiative transport in the hypersonic environment is modeled using the \texttt{MURP} (MUlti-fidelity Radiation Package) toolkit \cite{Jo2023a}.  This solver resolves the radiative transfer equation (RTE) for an absorbing, emitting, and non-scattering non-gray medium in multi-dimension. The present study employs a finite-volume approach to solve the transport equation in a two-dimensional axisymmetric configuration of interest. Spectral absorption and emission coefficients are calculated using a line-by-line method for a given thermo-chemical state, followed by model order reduction based on a multi-band opacity binning approach \cite{Johnston2018, Scoggins2013}. The radiation system covered in this study includes typical systems for the study of Earth entry environment: \emph{bound-bound}, \emph{bound-free} end \emph{free-free} transitions of N and O, as well as bound-bound systems of N$_2$, O$_2$, NO, and N$_2^+$ ranging from vacuum UV to near-infrared. In addition, the spectral properties of the ablation species are also considered, including the C, CO, C$_2$, and CN systems \cite{Johnston2009}. The considered spectral range spans 50 to 10,500 nm. Once the solution of RTE is obtained in terms of the spectral radiative intensity $I_\lambda^g$ at the given wavelength $\lambda$, the local volumetric radiative source term in the flow, $\Omega_{rad}^g$, is computed as

\begin{equation}\label{divQ}
    \Omega_{\mathrm{rad}}^g=\int_0^{\infty} \kappa_\lambda^g\left[4 \pi I_{b, \lambda}-\int_{4 \pi} I_\lambda^g d \theta\right] d \lambda,
\end{equation}

\noindent
where $\kappa_\lambda^g$, $I_{b,\lambda}$, and $\theta$ are the absorption coefficient, the Planck function, and the solid angle, respectively. The superscript $g$ denotes gas-phase relative physical variable. $\Omega^{g}_{\text{rad}}$ is communicated to the flow solver \texttt{HEGEL}. Over the TPS surface, the wall-directed radiative heat flux, $\omega_{\mathrm{rad}}^g$, can be calculated as:

\begin{equation}\label{heatRad}
    \omega^{g}_{\text{rad}} = \int_0^{\infty} \int_{2\pi} I_\lambda^g \Big|_{\vec{r}_w,\vec{s}} (\vec{s} \cdot \vec{n}_w) d\theta d\lambda,
\end{equation}

\noindent
where $I_\lambda^g \Big|_{\vec{r}_w,\vec{s}}$ denotes the spectral radiative intensity along a wall-directed ray $\vec{s}$ at the TPS surface $\vec{r}_w$. $\vec{n}_w$ is the unit normal vector at the wall. $\omega^{g}_{\text{rad}}$ is then communicated to the material solver \texttt{PATO} for the energy balance along the TPS surface.

\subsection{Thermo-Chemical Non-Equilibrium Flow}
%\subsection*{Thermo-Chemical Non-Equilibrium Flow}

Hypersonic plasma flows are simulated through the numerical solution of the non-equilibrium two-temperature reactive compressible Navier-Stokes system of equations, implemented in the \texttt{HEGEL} solver (High-fidelity tool for maGnEto-gas dynamics simulations) \cite{Munafo2020}. This numerical toolkit uses a finite volume discretization method on a multi-block structured spatial grid topology with a backward Euler implicit local time-stepping scheme for time integration towards a steady-state flow. This work uses a second-order MUSCL reconstruction scheme on a shock-tailored grid to accurately resolve the shock front and the boundary layer.

The set of chemical species used to compute the flows consists of the nominal air-11 species for Earth entry \cite{Park2001} with the addition of ablative products for carbon-based material and their ionic counterpart: that is, C, C$^+$, CO, CO$^+$, CO$_2$, CN, CN$^+$, C$_2$, and C$_3$. The corresponding kinetic mechanism and its rate coefficients are taken from the existing study on hypersonic flow and material response interactions \cite{Johnston2014}.

The volumetric radiative source term in Eq.~(\ref{divQ}) is injected into the total and the vibrational-electronic-electron energy equations \cite{Jo2023b}:

\begin{equation} 
    \frac{\partial}{\partial t} \left( \rho^g E^g \right) + \nabla \cdot \left( \rho^g H^g \vec{u}^g \right) + \sum_i \rho^g h_i^g \vec{U}_i^{g} = \nabla \cdot \left( \mathbf{\tau}^g \cdot \vec{u}^g \right) - \nabla \cdot \vec{q}^g + \Omega_{\text{rad}}^g,
\end{equation}

\begin{align}\label{eq-Energy}
    \frac{\partial}{\partial t} \left( \rho^g e_{\text{ev}}^g \right) 
    &+ \nabla \cdot \left( \rho^g e_{\text{ev}}^g \vec{u}^g \right) 
    + \sum_i \rho^g_i h_{\text{ev},i}^g \vec{U}_i^g \nonumber \\
    &= -p_{\text{e}}^g \nabla \cdot \vec{u}^g 
    - \nabla \cdot \vec{q}_{\text{ev}}^g 
    + \Omega_{\text{rad}}^g 
    + \Omega_{\text{ET}}^g 
    + \Omega_{\text{CE}}^g 
    + \Omega_{\text{VT}}^g,
\end{align} 

\noindent
where $t$, $\rho^g$, and $E^g$ are correspondingly the time, the density, and the total energy. $H^g$, $\vec{u}^g$, and $\tau^g$ are the total enthalpy, the velocity, and the stress tensor, respectively. $h_i^g$ and $\vec{U}_i^g$ are the enthalpy and diffusion velocity of the given species $i$. $\vec{q}^g$ denotes the conductive heat flux vector. $e_{\text{ev}}^g$ and $h_{\text{ev},i}^g$ are the mixture and the species vibrational-electronic-electron (subscript $\text{ev}$) energy and enthalpy, respectively. $p_{\text{e}}^g$ and $\vec{q}_{\text{ev}}^g$ are the free electron pressure and the conductive heat flux from the non-equilibrium energy content. The thermo-chemical non-equilibrium source terms, the electron-translational ($\Omega_{\text{ET}}^g$) and vibrational-translational ($\Omega_{\text{VT}}^g$) energy transfers, and the chemistry-electronic energy coupling ($\Omega_{\text{CE}}^g$) are modeled by following the work of Munaf\`o et al. \cite{Munafo2020}. The thermal energy transfers and chemical reactive source terms in the right-hand side, except for $\Omega_{\text{rad}}^g$, of Eq. (\ref{eq-Energy}) are calculated by using a physico-chemical library \texttt{PLATO} (PLAsmas in Thermodynamic nOn-equilibrium) \cite{Munafo2023}.

At the material surface, the mass and energy balance between the ablative TPS and the boundary layer is solved to ensure accurate coupling between the material response and the surrounding flow field \cite{Martin2017}. In the flow solver, the temperature and composition of the vehicle's surface are prescribed, while the material mass blowing rate is imposed through a normal blowing velocity at the interface between the material and the boundary layer \cite{Thompson2012}:

\begin{equation}\label{SurfEQ}
\begin{gathered}
    T_{\mathrm{h}}^g = T_{\mathrm{ev}}^g = T^s, \\
    y_i^g = y_i^s, \\
    \rho^g = \rho^g\left(T^s, \vec{\dot{m}}^s\right), \\
    \vec{u}^g = \vec{u}^g\left(\vec{\dot{m}}^s, \rho^g\right).
\end{gathered}
\end{equation}

\noindent
Here, $\vec{\dot{m}}^s$ represents the mass flux blown from the material surface into the boundary layer. The surface coupling of the equilibrium boundary conditions is governed by Dirichlet-type boundary conditions, as shown in Eq.~(\ref{SurfEQ}). $T^s$ and $y_i$ denote the material temperature (superscript $s$) and the mass fraction of the species, respectively. These conditions are dynamically updated in response to changes in the material surface state, which is achieved using the proposed multi-solver coupled approach.

\subsection{Material Response}
To reduce the number of species considered within the material and in the boundary layer, the TPS material in this study is assumed to be non-pyrolyzing and characterized by relatively low porosity. This assumption allows for neglecting the volumetric transport of boundary layer gases within the material. %, UNCLEAR WHAT THIS MEANS except in cases of finite-rate ablation responses, addressed in the following paragraphs.
Furthermore, it excludes the impact of pyrolyzed species such as H, H$_2$O, and CH$_4$ in the boundary layer \cite{Martin2011}. This simplification is based on the assumption that the effects of these pyrolyzed species on convective and radiative heat transfer are minimal in high-altitude entry conditions \cite{Johnston2014}, which are the focus of this study.

One of the goals of this work is to assess the validity of the equilibrium assumption at the material surface by comparing it with finite-rate ablation processes. The equilibrium surface response assumption is examined at flight conditions corresponding to the peak heating phase of the FIRE II entry case \cite{Johnston2009}. In this high-density regime, the equilibrium assumption in the boundary layer is considered reasonable because of the relatively higher free-stream density. 

To validate this assumption, we apply equilibrium boundary conditions (BCs) in conjunction with the blowing boundary conditions derived from the B' methodology \cite{Meurisse2018}. These are compared with results obtained using the Park-5 ablation response model to highlight the differences between equilibrium and finite-rate ablation dynamics. This comparison is crucial for understanding the limits of the equilibrium hypothesis, particularly in high-altitude entry scenarios where non-equilibrium effects may become significant.

The mass balance of the species on the surface of the nonpyrolyzing TPS material is expressed as~\cite{Martin2011}:

\begin{equation}\label{EqMassSurf}
    \left(\rho_i^g \vec{u}^g-\vec{J}_i^g\right) \cdot \vec{n}_w=\dot{s}_i^s,
\end{equation}

\noindent
where $\vec{J}_i^g$ denotes the species diffusion flux. The surface heterogeneous species production term, $\dot{s}_i^s$, is explicitly calculated using the finite-rate Park-5 model \cite{Johnston2014}. In doing so, Eq. (\ref{EqMassSurf}) can be prescribed directly as a non-homogeneous Neumann boundary condition for the mass conservation equation of species inside the material, which is required to explicitly account for the porous nature of the TPS. If the material transport parameters (\emph{e.g.}, permeability) are low enough, a steady state for chemical species concentration is quickly reached at the surface. The updated local mass fraction of the species over the interface $y_i^s$ is then communicated back over the surface to the CFD. The total blowing rate is computed by summing the surface heterogeneous source term over all the sets of species:

\begin{equation}
  \vec{\dot{m}}^s \cdot \vec{n}_w=\sum_i \dot{s}_i^s.  
\end{equation}

In contrast, the equilibrium wall assumption simplifies the ablation process by treating it as a surface-specific phenomenon, eliminating the need to account for the in-depth transport of gas species within the TPS material. This approach has been shown to significantly reduce computational time and enhance the stability of the coupling methodology \cite{LeMaout2023}. The wall composition can be determined directly through a Gibbs free energy minimization algorithm \cite{Scoggins2017}, which utilizes the thermodynamic properties of condensed graphite and the elemental composition of the hypersonic boundary layer. This method enables the direct prescription of the species mass fractions at the surface, $y_i^s$, as Dirichlet boundary conditions. The conservation of the carbon elemental mass fraction, $z_{\text{C}}$, on the wall leads to:
\begin{equation}\label{carbonBalance}
    \left(z_{\text{C}} \rho^g \vec{u}^g-\vec{J}_{\text{C}}^g\right) \cdot \vec{n}_w=\vec{\dot{m}}^s z_{\text{C}} \cdot \vec{n}_w,
\end{equation}

\noindent
where the diffusion flux of an element $k$ on the TPS surface (denoted $J_k^g$) is calculated using boundary layer scaling laws \cite{Schlichting2017}:

\begin{equation}\label{bprimeEq}
    \vec{J}_k^g \cdot \vec{n}_w =\rho_e^g u_e^g C_h\left(z_{k, e}^g-z_k^s\right).
\end{equation}

\begin{figure}[h!]
    \centering
    \includegraphics[clip,bb=0 75 880 375,width=1.0\textwidth]{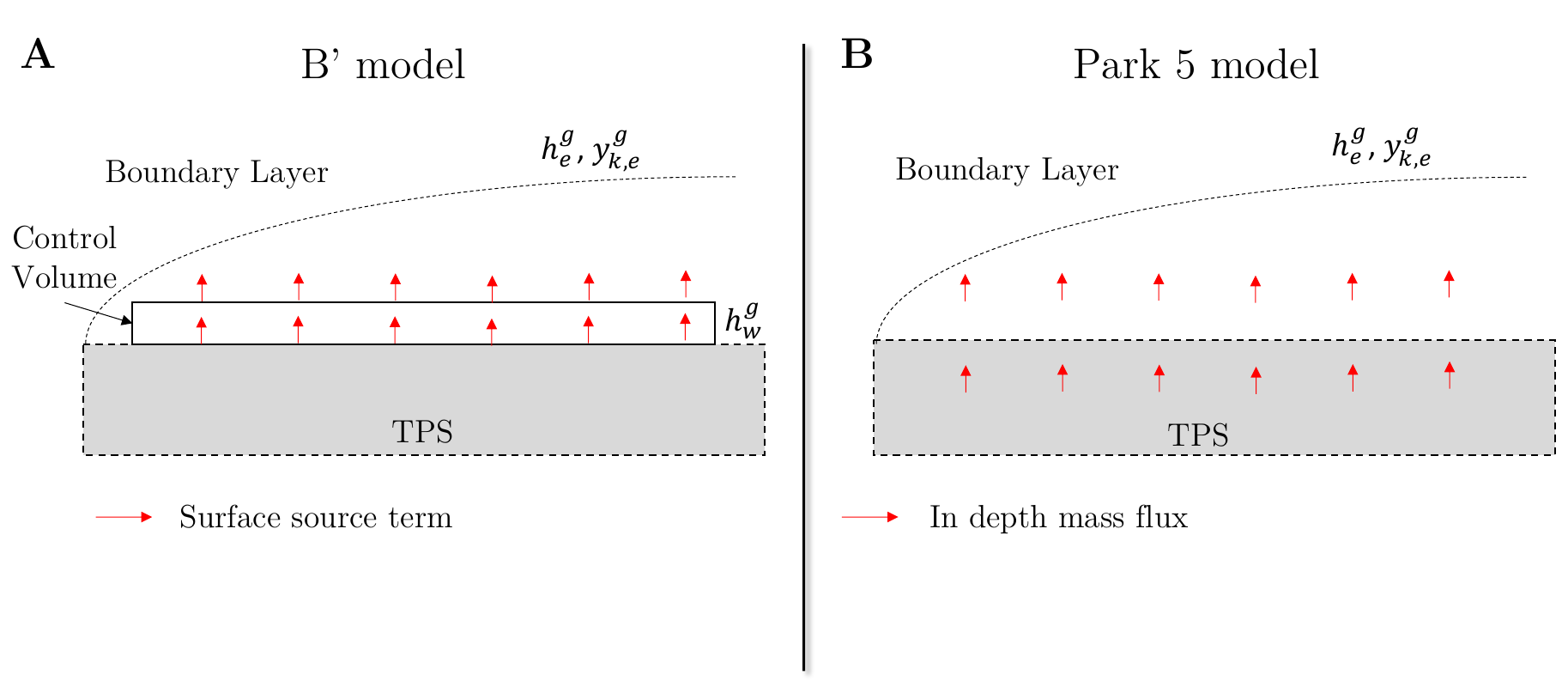}
    \caption{Schematic of the two ablation models investigated in this work. (A) B’ equilibrium ablation model. (B) Finite-rate Park-5 model.}
    \label{AblationModels}
\end{figure}

\noindent
In Eq. (\ref{bprimeEq}), the subscript $e$ refers to boundary layer edge evaluated properties. The Stanton number $C_h$ is computed according to the heat flux prescribed at the wall \cite{Meurisse2018, Martin2014}. Injecting Eq. (\ref{bprimeEq}) into the elemental mass conservation of carbon in Eq. (\ref{carbonBalance}) and knowing the elemental composition of the wall through the Gibbs free energy minimization allows one to compute the blowing rate $\vec{\dot{m}}^s\cdot \vec{n}_w$ for the $B’$ boundary condition and communicating it back to the flow solver. An overview of the two types of boundary conditions for a charring material surface is given in Fig. \ref{AblationModels}.

% Figure-2 in MS Word

The surface energy balance at the TPS boundary allows for the calculation of the total energy flux transferred from the boundary layer to the material, denoted as $\vec{F}^g$. This flux is decomposed into two components: the incoming convective-diffusive heat flux and the radiative heat flux at the surface, as expressed in Eq.~(\ref{heatRad}). The surface energy balance is represented by a non-homogeneous Neumann boundary condition applied to the material temperature equation, given by the following relation \cite{Johnston2014}:

\begin{equation}\label{BC_energy}
    -\underline{K} \nabla T^s \cdot \vec{n}_w=\vec{F}^g \cdot \vec{n}_w-\epsilon \sigma\left[\left(T^s\right)^4-T_{\infty}^4\right]-\vec{\dot{m}}^s \cdot \vec{n}_w \left(h_{\text{C}}^s-h_w^g\right),
\end{equation}

\noindent
where $\underline{K}$, $\epsilon$, and $\sigma$ are correspondingly the thermal conductivity, the TPS emissivity, and the Stefan-Boltzmann constant. At the material surface, the blackbody radiation emission law is applied as the primary energy dissipation mechanism at the steady state. Furthermore, the effect of ablative surface blowing is accounted for, as described in \cite{Meurisse2018}.

The surface energy balance at the material boundary introduces several non-linearities, as shown in Eq.~(\ref{BC_energy}). Typically, these are implemented using a linearized discrete model, which computes the radiative energy contribution based on previous time step values of pressure and temperature. However, in this work, a fixed point iteration approach is employed, utilizing the implicit coupling capabilities of the \texttt{preCICE} library. Sub-iterations between the material and CFD solvers are performed until convergence of the exchanged boundary data is reached.

\section{Results}\label{Sec:Results}
\subsection{Coupled Flow-Ablation Interaction}\label{CoupledAbl}
\subsubsection{Baseline Simulations}\label{Sec:Baseline}
%To investigate the interactions between radiative transfer and charring material, the simulation case uses the free-stream condition corresponding to the peak heating of the FIRE II trajectory, which has been demonstrated to raise a large amount of radiative heat flux in past studies \cite{Johnston2014, Jo2020}. Figure \ref{Baseline} visually describes the baseline flow field with the free-stream condition.\\

\begin{figure}[h!]\label{resStag}
    \centering
    \includegraphics[width=1.0\textwidth]{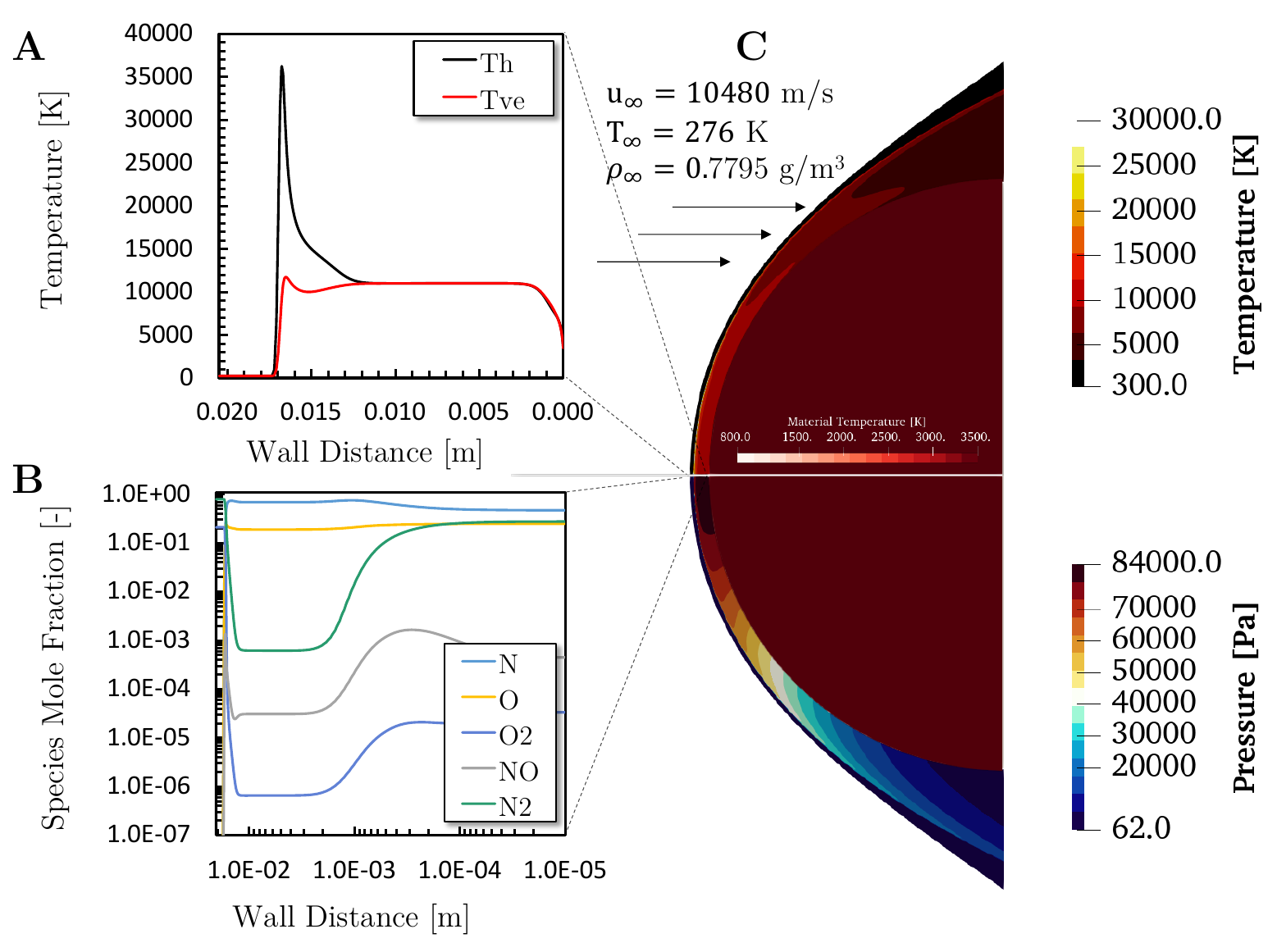}
    \caption{Baseline flow field for an isothermal surface (and material) at $T=\qty{3500}{\kelvin}$ with non-catalytic wall boundary conditions. (A) Temperature profile along the stagnation line. (B) Mole fraction variation of the main species along the stagnation line. (C) Contour plot of temperature and pressure around the blunt body.}
    \label{Baseline}
\end{figure}

This study examines the interaction between radiative transfer and material response under flight conditions corresponding to the peak heating phase of the FIRE II trajectory. At these conditions, radiative heating becomes comparable to convective heating \cite{Johnston2014, Jo2020}. Figure \ref{Baseline} presents the baseline flow field under these free-stream conditions.

This figure displays flow calculations for a non-catalytic, isothermal wall with a surface temperature of $T_s = 3500$~K applied to the TPS surface. The stagnation line profiles in Figs. \ref{Baseline}-A and \ref{Baseline}-B reveal the presence of thermo-chemical non-equilibrium effects within the shock layer, while the boundary layer remains near equilibrium in terms of gas temperature and pressure. Due to the elevated temperature in the boundary layer, the free-stream molecules in the post-shock region near the stagnation point are highly dissociated, assuming no surface concentration is enforced on the vehicle's surface.

\begin{figure}[h!]
    \centering
    \includegraphics[width=0.9\textwidth]{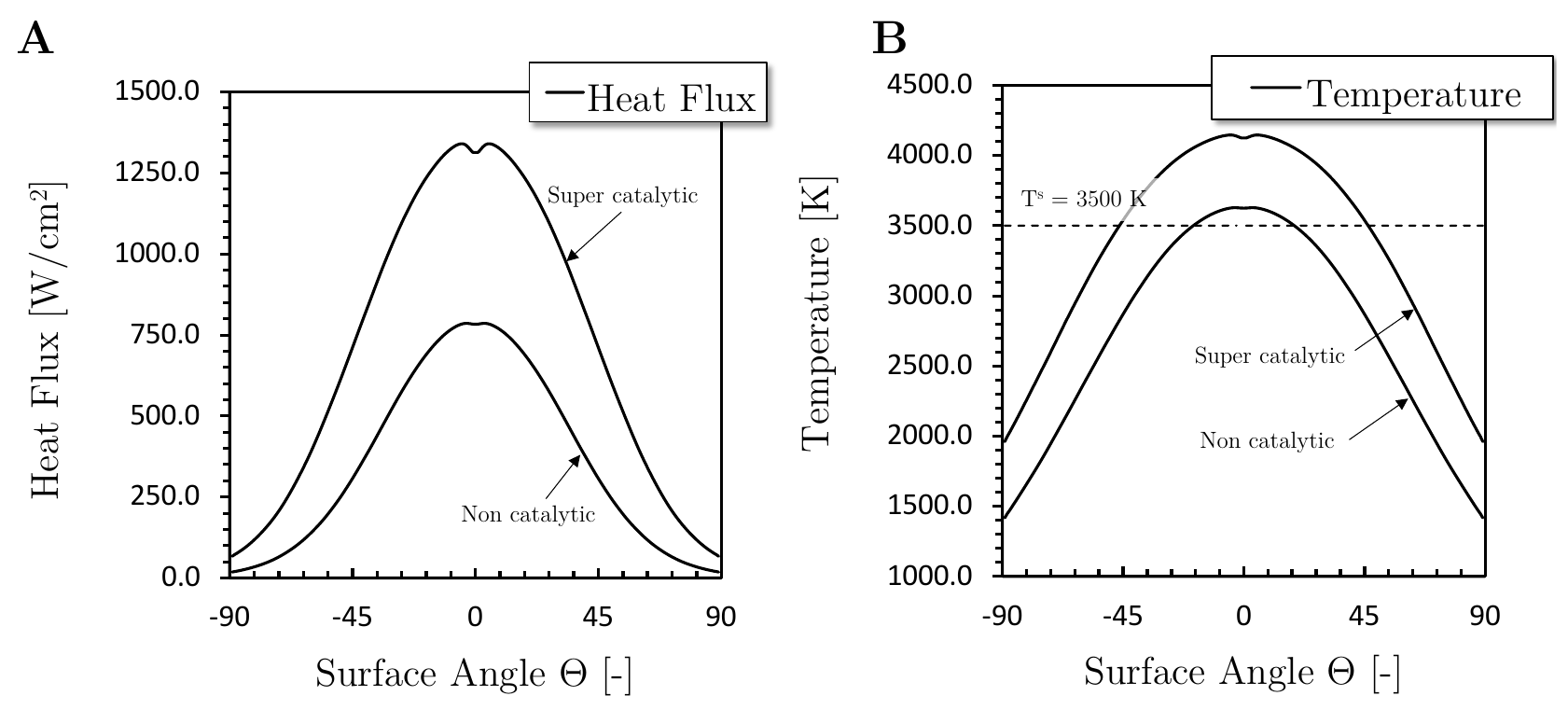}
    \caption{Profiles of (A) wall heat flux and (B) wall temperature at theoretical radiative equilibrium for the baseline condition.}
    \label{BaselineWall}
\end{figure}

 In the absence of surface blowing, the surface heat flux, shown in Fig.~\ref{BaselineWall}-A, is significantly high near the stagnation point, as expected for high-speed atmospheric entry. This remains true even under a non-catalytic boundary condition, where no chemical reactions contribute to the heat flux, excluding radiative heat transfer. However, as shown in Fig.~\ref{BaselineWall}-A, when a super-catalytic boundary condition is assumed (\emph{i.e.}, the free-stream species mass fractions are fully recovered at the TPS surface), the chemical contribution increases the heat flux by up to 70\% near the stagnation point. This increase is due to the highly dissociated state of the gas-phase species in the boundary layer, as illustrated in Fig.~\ref{Baseline}-B. 

The corresponding theoretical wall temperature, depicted in Fig.~\ref{BaselineWall}-B, exceeds the prescribed 3,500 K under both non-catalytic and super-catalytic simulations if blackbody radiation is assumed to be the sole heat dissipation mechanism. In an actual flight scenario, additional heat dissipation mechanisms would prevent the surface from reaching such high temperatures, and the TPS material would likely consist of more than just carbon \cite{Uyanna2020}. However, the study of material response in such aerothermal environments has previously been explored \cite{Candler2012}, and remains of significant interest in high-speed atmospheric entry contexts. %Consequently, in this study, both non-catalytic and super-catalytic boundary conditions are considered as analogs for the equilibrium and finite-rate ablation responses, which are discussed in the following sections.

\subsubsection{Investigation of Surface Ablation Response}
\begin{figure}[h!]
    \centering
    \includegraphics[width=1.0\textwidth]{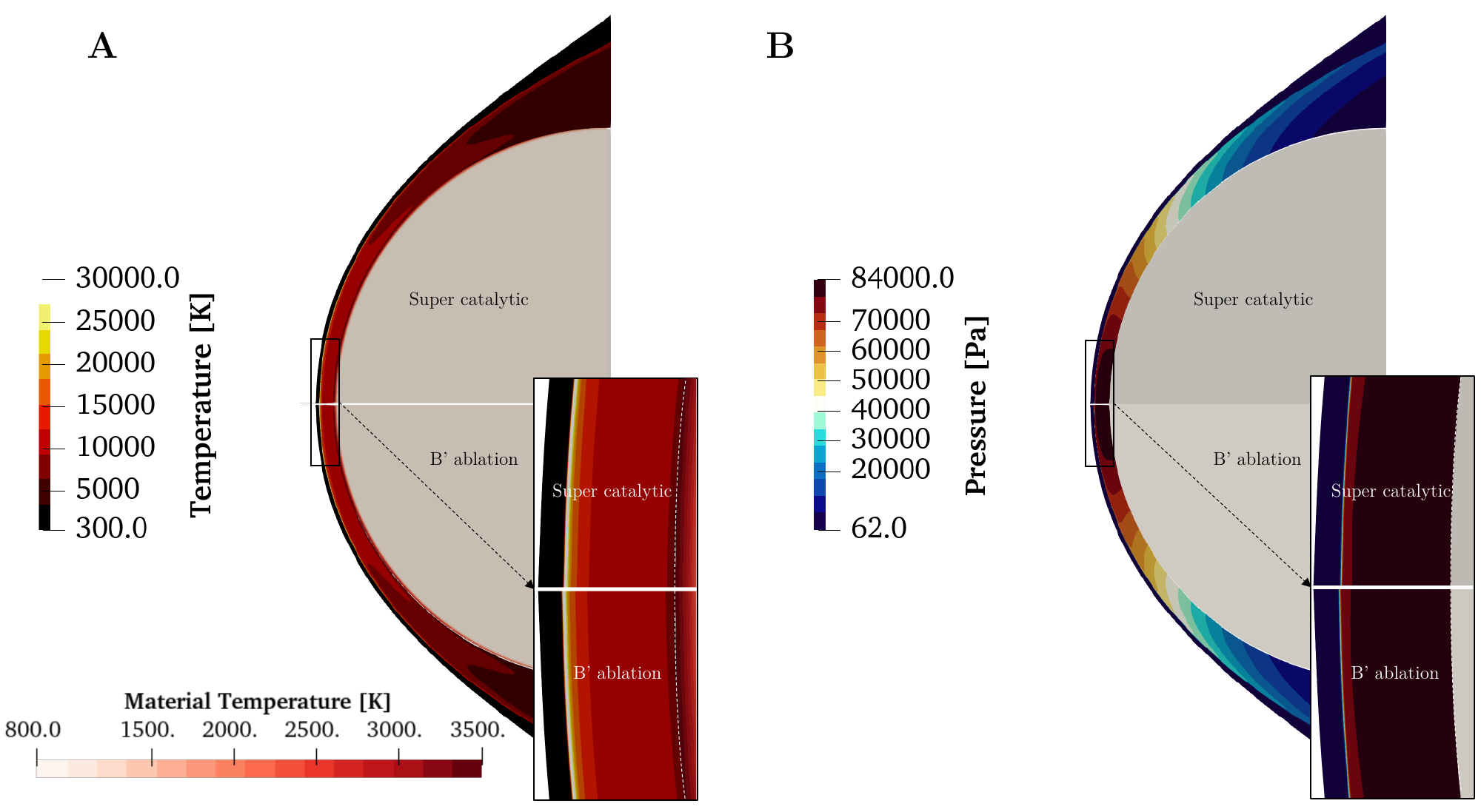}
    \caption{Front body simulation comparisons for the 1643s trajectory point of FIRE II between a super-catalytic and an equilibrium ablation TPS surface boundary condition. (A) $T$ and (B) $p$ contours. Insert refers to the solution in a region zoomed around the stagnation line.}
    \label{SC_Bp_Contour}
\end{figure}

The solution presented in Sec. \ref{Sec:Baseline} is used to compute the material response of an ablative surface. Unlike previous simulations, the surface temperature is not prescribed but determined as the simulation progresses until the surface reaches a steady-state condition. Figure \ref{SC_Bp_Contour} provides an overview of the computed solution field under the B’ boundary condition, with comparisons made against the results from the super-catalytic case.

% Figure-5 in MS Word

%In Fig. \ref{SC_Bp_Contour}, the flight conditions corresponding to the investigated trajectory point impose fluid dynamic conditions that the vehicle wall blowing rate has minimal impact on the shock front location. However, near the stagnation point, where the surface blowing rate is highest, the influx of ablative species into the boundary layer causes a slight shift in the shock location (see Fig. \ref{SC_Bp_Contour}-A insert) when compared to the non-ablative super-catalytic case. A more detailed analysis of the stagnation line temperature and composition, presented in Fig.~\ref{SC_Bp_Stagnation}-A and \ref{SC_Bp_Stagnation}-B, reveals the differences between the two cases, particularly with regard to the gas-phase composition.\\

\begin{figure}[h!]
    \centering
    \includegraphics[width=1.0\textwidth]{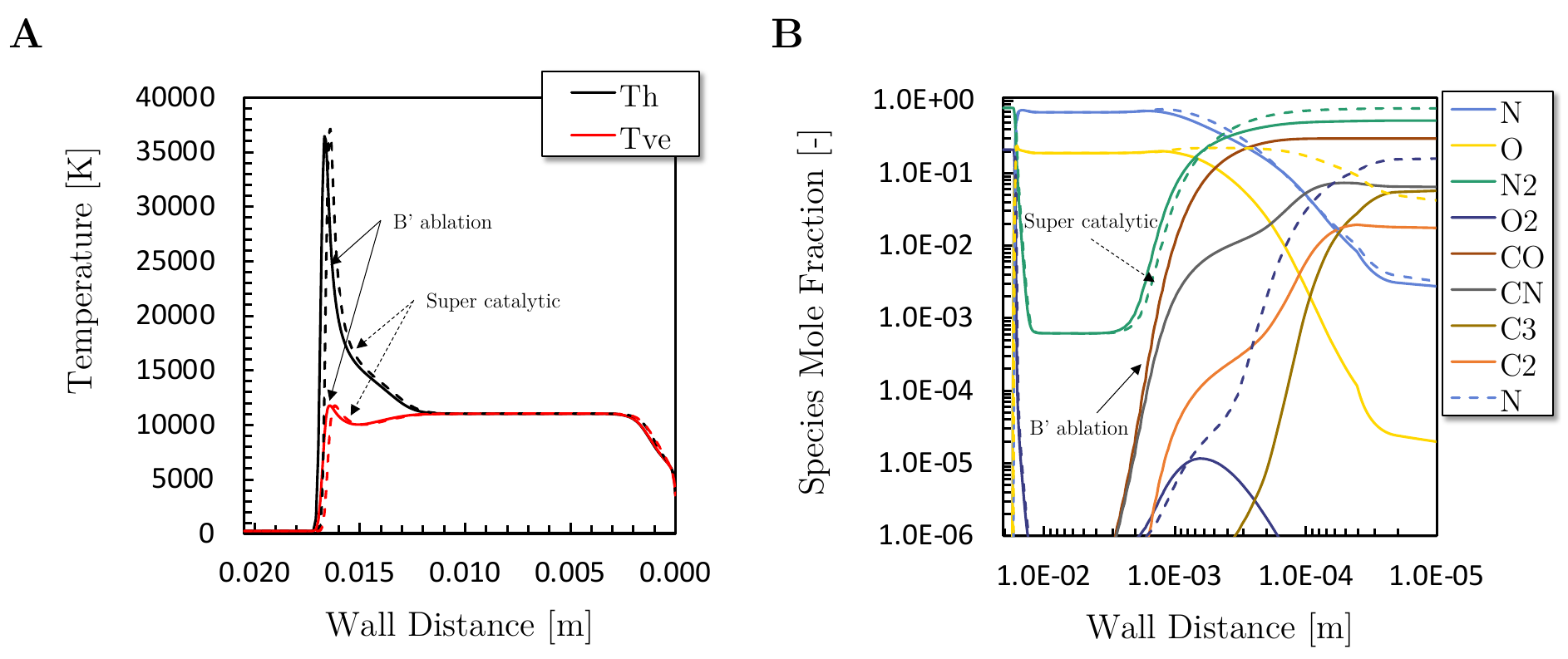}
    \caption{Stagnation line profiles for the investigated entry case for the super-catalytic and B’ ablation models. (A) Temperatures and (B) main boundary layer species mole fraction distributions. In all plots, the dashed lines are for the super-catalytic case, while the solid lines refer to the B’ ablation model.}
    \label{SC_Bp_Stagnation}
\end{figure}

In Fig. \ref{SC_Bp_Contour}, the flight conditions at the investigated trajectory point create aerothermal conditions such that the vehicle wall blowing rate has minimal influence on the shock front location. However, near the stagnation point, where the surface blowing rate is highest, the influx of ablative species into the boundary layer results in a slight shift of the shock location (see Fig.~\ref{SC_Bp_Contour}-A insert) compared to the non-ablative super-catalytic case. A closer analysis of the stagnation line temperature and composition, shown in Figs.~\ref{SC_Bp_Stagnation}-A and \ref{SC_Bp_Stagnation}-B, highlights the differences between the two cases, particularly in terms of gas-phase composition.

% Figure-6 in MS Word

Because the ablation products from carbon-based materials introduce foreign species into the flow domain, the composition of the shock layer undergoes significant changes when the super catalytic approach is switched from B'. In the B’ ablation model, CO, CN, and C$_3$ are introduced into the boundary layer as ablation products, while O and O$_2$ are entirely consumed at the surface, as shown in Fig.~\ref{SC_Bp_Stagnation}-B. Additionally, a portion of the nitrogen (N) reaching the surface is consumed in nitridation processes, resulting in lower mole fractions of N$_2$ and N at the stagnation point compared to the super-catalytic case. 

In regions where the heating is less intense, such as near the shoulder of the vehicle, the B’ ablation model predicts primarily carbon oxidation, leading to the release of only CO molecules into the boundary layer. The effect of the gas kinetics model is particularly noticeable for carbon-based species, as seen in Fig.~\ref{SC_Bp_Stagnation}-B. Species with relatively low chemical reactivity, such as CO, exhibit a monotonous diffusion through the boundary layer, while CN, C$_3$, and C$_2$ display non-monotonous behavior. The maximum CN concentration occurs a few micrometers away from the wall, decreasing rapidly. Similar trends are observed for C$_3$ and C$_2$, which react quickly near the wall with species originating from the free-stream.

Because existing coupling studies employed either the wall equilibrium hypothesis or the finite-rate ablation model, a quantification of the introduced differences is investigated in this part. The Park-5 ablation model is directly compared with the previous results obtained using a B’ approach to evaluate potential deviations due to the choice of ablation models at the given trajectory point. The finite-rate surface parameters used in this study are taken from the previous study~\cite{Johnston2012} and are summarized in Table \ref{table:Park-5} of Appendix.
The results are presented in Fig. \ref{Park5_Bp_Contour}. As observed in the comparison between the equilibrium ablation B model with the super-catalytic wall boundary, the shock location is only slightly affected by the variation of the ablation models (c.f. Fig. \ref{Park5_Bp_Contour}-A and its insert). However, noticeable differences are observed when investigating the surface data exchanged between the flow solver and the material solver closer. In particular, an increase of 5\% of the surface heat flux is observed as shown in Fig. \ref{Park5_Bp_Contour}-B for the Park-5 ablation model at steady state compared to the B’ approach. The latter remains below the value observed for the super-catalytic wall boundary conditions. Despite the rise of the heat flux, the wall temperature remains lower for the Park-5 model when compared with the B’ result, with a difference of around 100 K. These results are similar to those of existing studies that compared different material response models \cite{Candler2012, Johnston2014}.

% Figure-7 in MS Word
\begin{figure}[h!]
    \centering
    \includegraphics[width=1.0\textwidth]{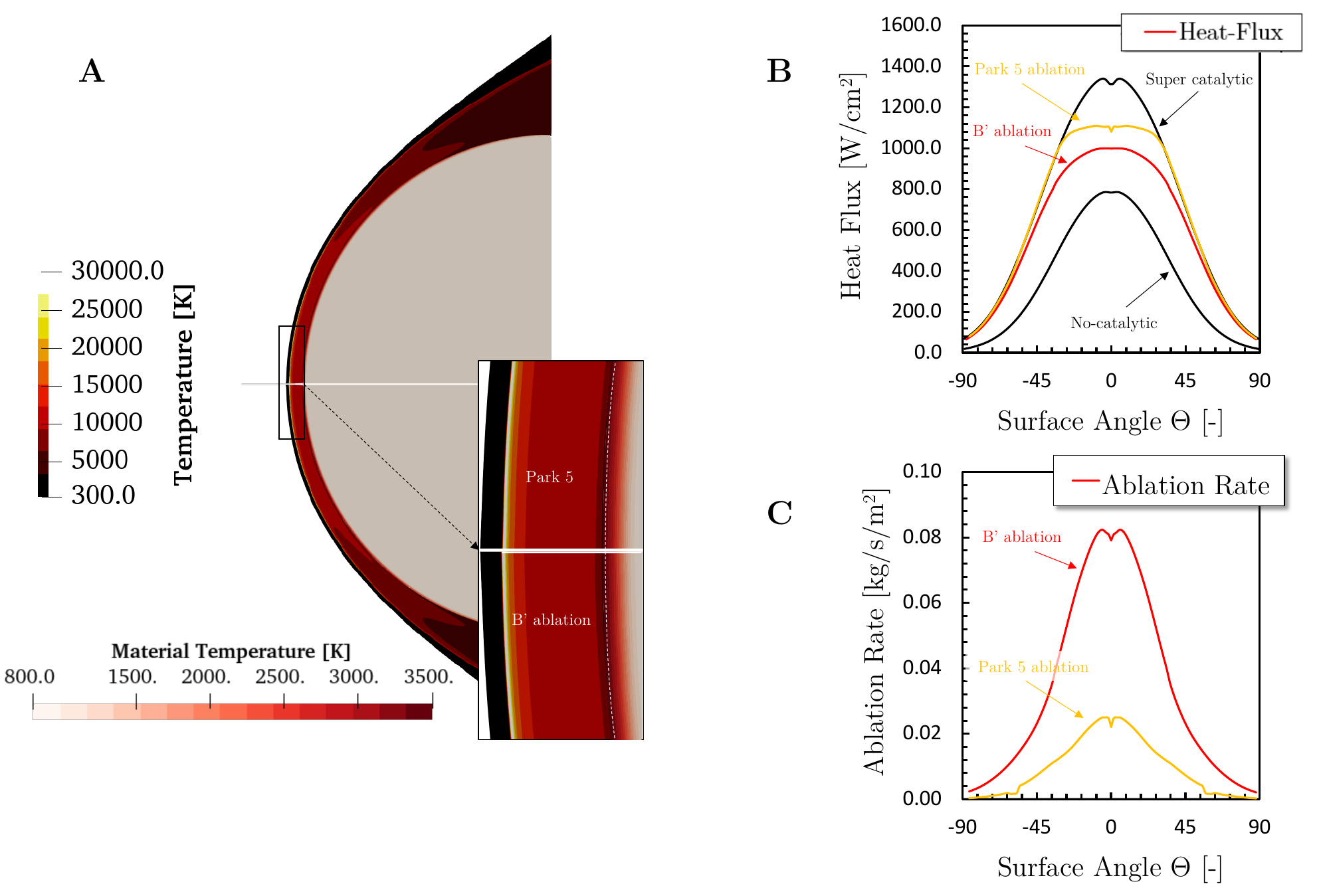}
    \caption{Comparison of equilibrium B’ ablation model with finite-rate Park-5 ablation approach. (A) Shock configuration around a blunt body. (B) Comparison of surface heat flux around the sphere for non-ablative cases (non-catalytic and super-catalytic) and ablative cases (B’ and Park-5 model). (C) Blown mass flux around the blunt body surface.}
    \label{Park5_Bp_Contour}
\end{figure}

\begin{figure}[h!]
    \centering
    \includegraphics[width=1.0\textwidth]{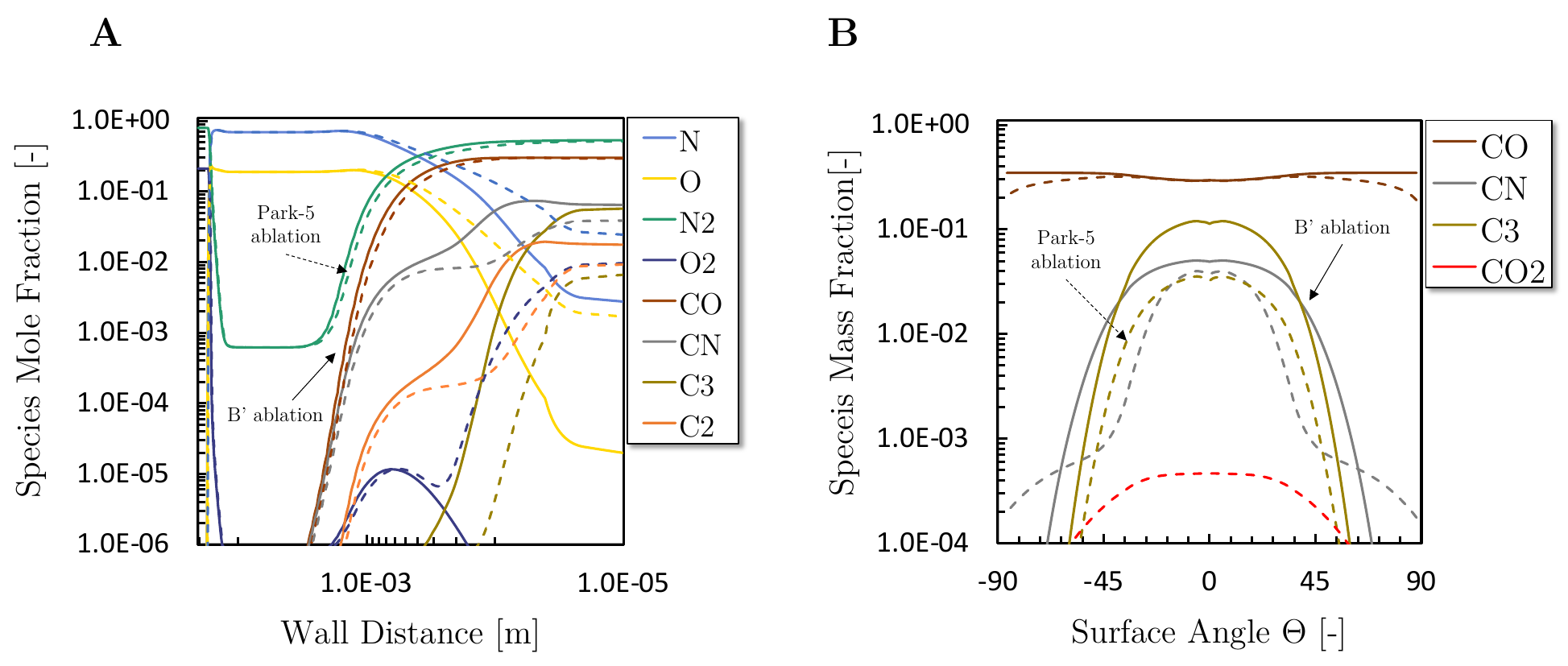}
    \caption{Comparisons of the composition profiles from the B’ ablation model (solid lines) with the Park-5 finite-rate ablation approach (dashed lines). (A) Stagnation line composition of the main species of the mixture. (B) Ablation product compositions over the blunt body surface.}
    \label{Park5_Bp_Stag_Wall}
\end{figure}

The lower surface temperature, as well as the kinetic model used in this study, lead to significantly lower values of carbon surface blowing rate (c.f. Fig. \ref{Park5_Bp_Contour}-C) for finite-rate model, similar to the previous study on carbon ablation in a high-speed Earth entry trajectory \cite{Candler2012}. Along the stagnation line, the difference between the B’ equilibrium blowing rate and the Park model becomes more critical, almost a factor three discrepancy between the models. When observing the composition of the gas-phase along the stagnation line, as shown in Fig. \ref{Park5_Bp_Stag_Wall}-A, the main difference between the two cases is the presence of O and O$_2$ near the wall for the Park-5 model, contrary to the B’ approach, which inherently assumes complete oxygen consumption.

% Figure-8 in MS Word

For the finite-rate model, the stagnation line and surface composition profiles show that the ablation products (CO, CN, and C$_3$) are blown in higher quantity for the B’ equilibrium model (c.f. Fig. \ref{Park5_Bp_Stag_Wall}-B). As described in the previous subsection, the wall composition is enforced using a Dirichlet boundary condition over the vehicle boundary according to the equilibrium computation. For this reason, no CO$_2$ formation is predicted with the given temperature for the B’ ablation model. On the contrary, even if no CO$_2$ is initially expected from the carbon ablation in the Park-5 model (c.f. Table \ref{table:Park-5}), the non-flux condition does not prevent the diffusive transport of CO$_2$ from the boundary layer on the wall, as shown in Fig.~\ref{Park5_Bp_Stag_Wall}-B.

By comparing the methodology of B' and the Park-5 ablation model for the material response, it is observed that the assumption of the surface in equilibrium state is not too far from the wall composition predicted by the finite-rate model, although it has been demonstrated that not all the incoming flux of O$_2$ participated in oxidation in the latter case. However, the large discrepancy calculated for the wall blowing rate indicates that most of the uncertainty of the B’ model lies in estimating the Stanton number in Eq. (\ref{Park5_Bp_Wall_Alpha}) to calculate the amount of mass ejected from the vehicle surface into the boundary layer. To investigate its influence, a sensitivity factor $\alpha$ is introduced to the computation of the boundary layer properties in this work, similar to the existing physics-based corrections parameters available in the literature \cite{Martin2011}:

\begin{equation}
    \vec{\dot{m}}^s \cdot \vec{n}_w=\alpha\left(\rho_e^g u_e^g C_h B^{\prime}\right)
\end{equation}

Introducing this coefficient allows one to linearly correct the blowing rate according to the Stanton number. Depending on the value of $\alpha$, it is anticipated that the blowing rate of the finite-rate model is expected to be recovered by the B' approach. A factor $\alpha=1$ corresponding to the nominal computation of the blowing rate is denoted by \emph{nominal} hereafter.

\subsubsection{Characterization of Surface State Sensitivity to Stanton Number}
\begin{figure}
    \centering
    \includegraphics[width=0.96\textwidth]{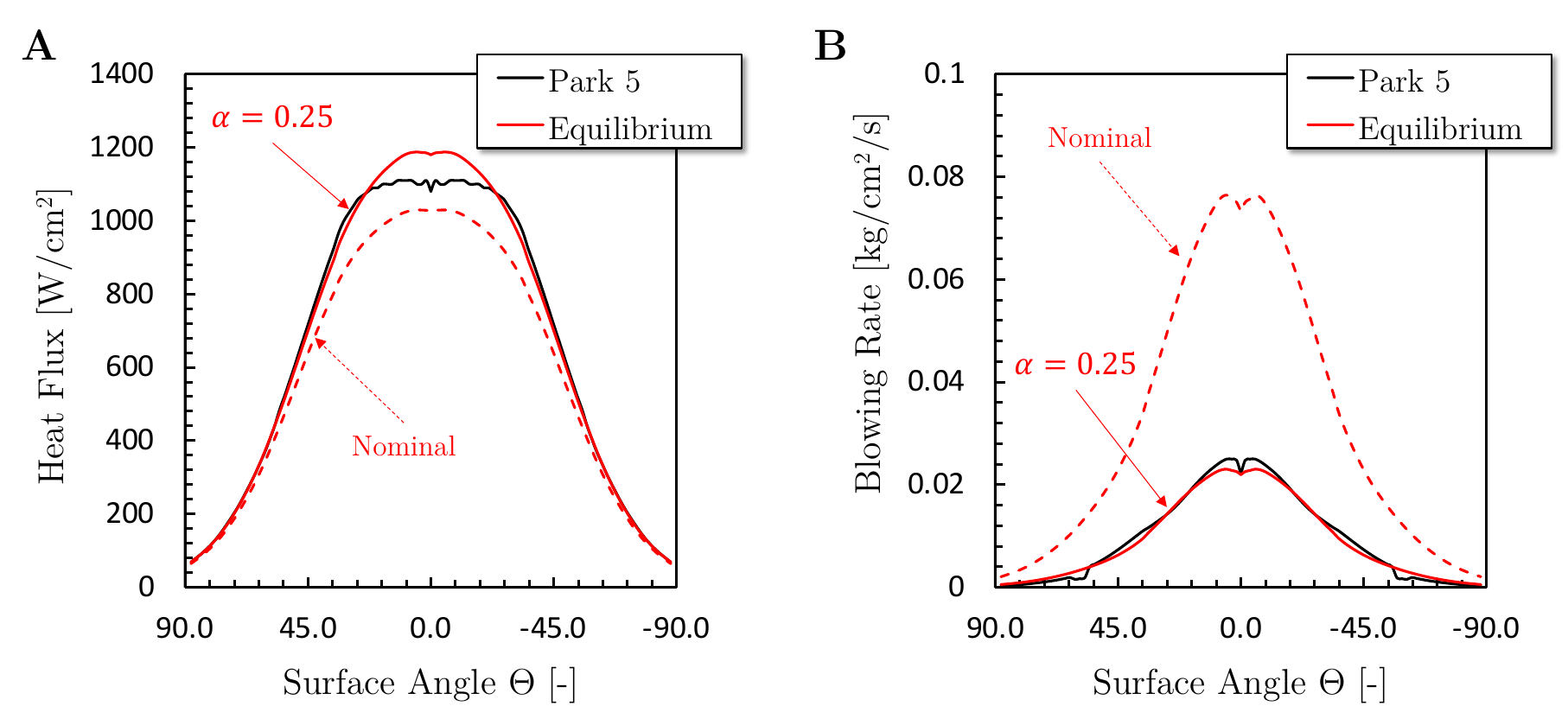}
    \caption{Comparisons of (A) the equilibrium model heat flux and (B) mass blowing rate with the finite-rate approach for a correction parameter $\alpha=0.25$.}
    \label{Park5_Bp_Wall_Alpha}
\end{figure}

The influence of the sensitivity parameter $\alpha$ is explored first without considering the radiative transfers. Figure \ref{Park5_Bp_Wall_Alpha}  shows the influence of variation of this parameter on the vehicle's surface response once the wall temperature reaches a steady state.

% Figure-9 in MS Word

Figure \ref{Park5_Bp_Wall_Alpha} demonstrates that, under \emph{a priori} calibration, the equilibrium ablation model can reproduce the thermodynamic state of the wall closer to the Park-5 carbon ablation model under the investigated flight condition both in terms of blowing rate and heat flux. Observed differences between the models persist around the stagnation point and indicate that the stagnation area is probably not fully recoverable with a pure equilibrium-based thermodynamic model.  However, from a numerical point of view, it is more interesting now to impose the Dirichlet boundary condition brought about by the equilibrium model in terms of numerical stability and steady-state convergence rate. Furthermore, the effect of finite-rate ablation appears to be contained in Figure \ref{Park5_Bp_Wall_Alpha} between the deviations observed for the equilibrium model for the $\alpha$ parameter taken between 1 and 0.25, which implies that the equilibrium model can be used to estimate an envelope of the ablative response. 

In light of the findings of this subsection, the ablation response considered in the next part will be restricted to the equilibrium B' model, which allows a faster convergence of the coupled framework towards a steady state. The parameters $\alpha$ are kept to study the potential modification of the solution brought about by a different ablation model under the hypothesis of equilibrium wall when coupling the flow material framework to the radiative solver.

\section{Coupled Flow-Ablation-Radiation Interactions}\label{CoupledRes}
Radiative transfers have now been added to the previous simulations. The solver \texttt{MURP} and \texttt{HEGEL} share the same computational grid, so the mesh refinement around the shock is used in the flow field and radiative response computation.

\subsection{Influence of Coupled Radiation Interaction on Equilibrium Ablation surface}
\begin{figure}
    \centering
    \includegraphics[width=0.96\textwidth]{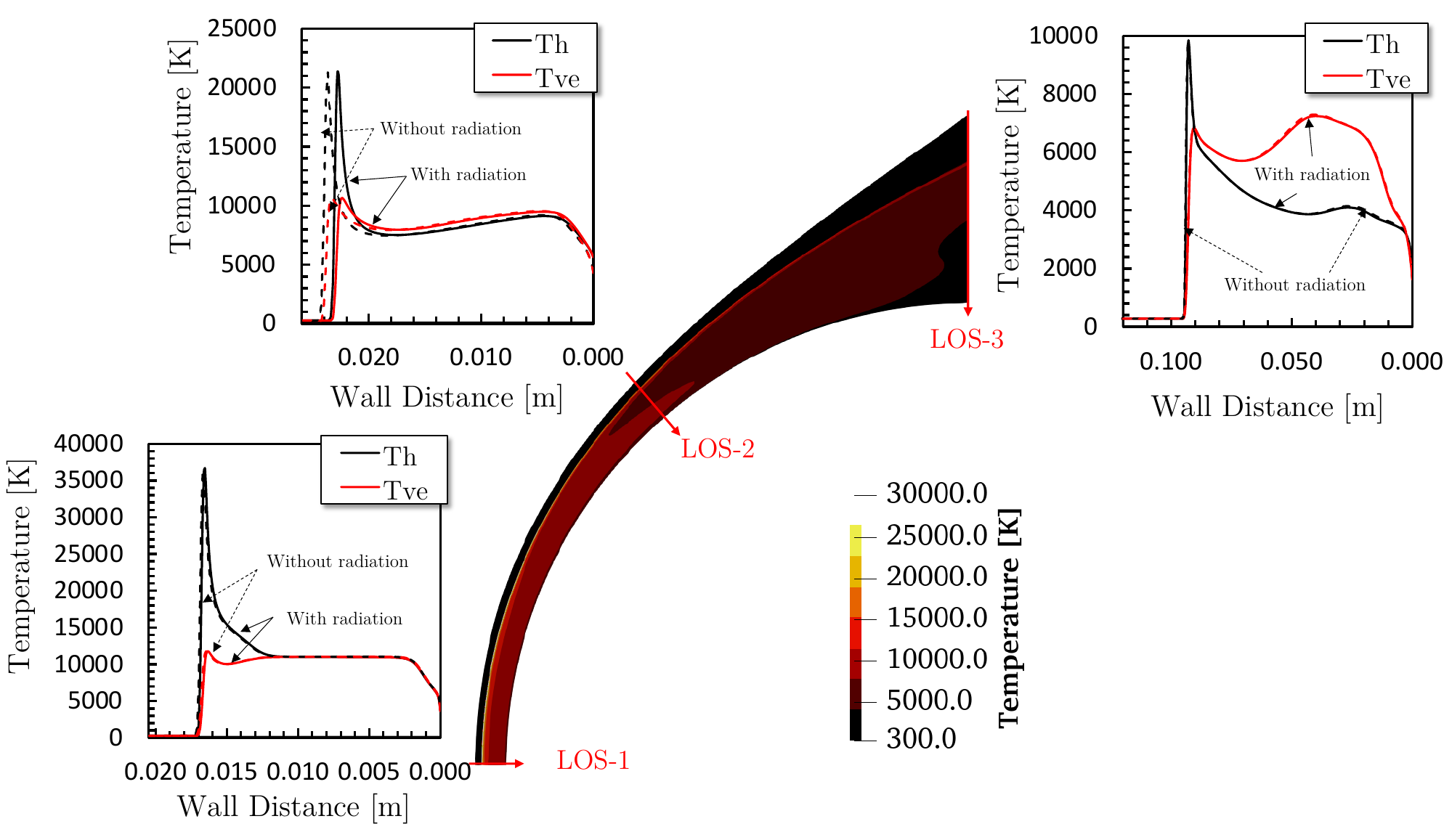}
    \caption{Comparisons of the temperature profiles with the B’ equilibrium ablation model with and without considering the radiative transfer. The temperature profiles are plotted along the three lines of sight (LOS).}
    \label{Bp_RadiationCoupling_Contour}
\end{figure}

A comparison of the results obtained with and without considering radiative transfers for the B' ablation method is provided in Fig.\ref{Bp_RadiationCoupling_Contour}. It can be observed that considering the radiative cooling decreases the boundary layer temperature around the surface, according to the previous study \cite{Jo2023b}, by enabling the energy loss of the computational domain. This additional energy dissipation in the high-temperature region in non-equilibrium slightly affects the shock configuration that is shifted closer to the vehicle surface.

% Figure-10 in MS Word

\begin{figure}[h!]\label{fig11}
    \centering
    \includegraphics[width=1.0\textwidth]{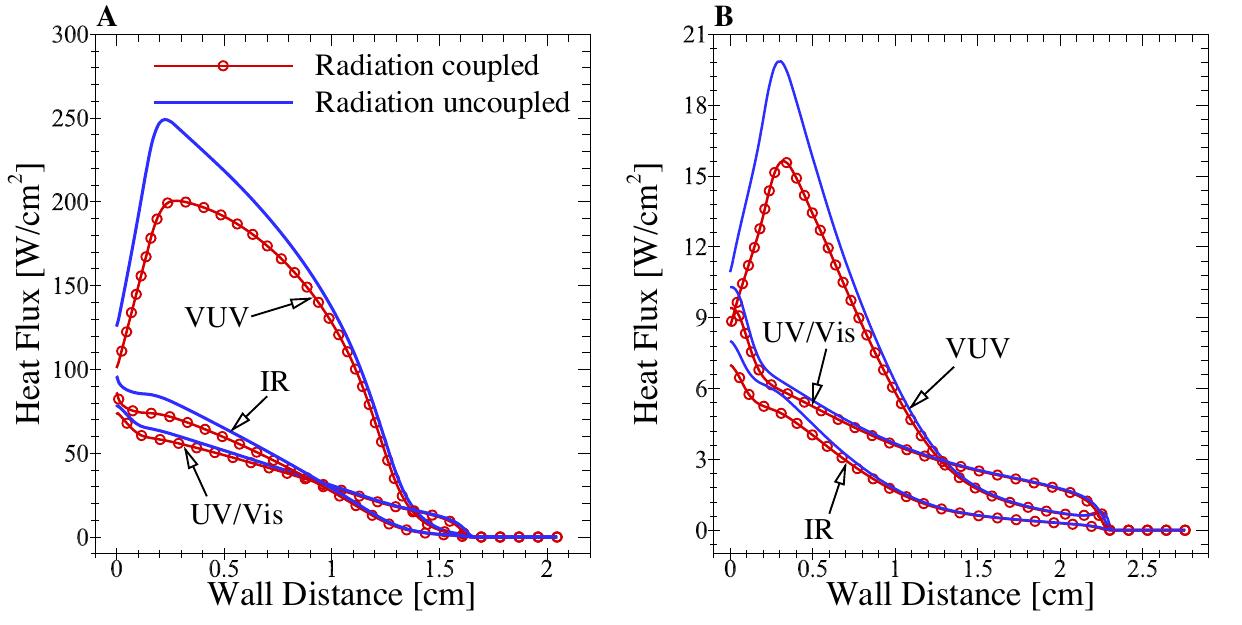}
    \caption{Influence of the radiation coupling on the radiative heat flux profiles. (A) LOS-1. (B) LOS-2.}
    \label{Qrad_CouplingEffect}
\end{figure}

The influence of the radiation coupling is further quantitatively analyzed by comparing the radiative heat flux profiles along the LOS-1 and LOS-2, as shown in Fig.~\ref{Qrad_CouplingEffect}. In both the stagnation line and the off-stagnation, the heat flux in the vacuum-ultraviolet (VUV) wavelength range undergoes the most prominent impact due to the radiation coupling, resulting in the most sensitive decreases of the heat flux compared to the ultraviolet and visible(UV/Vis) and infrared(IR) ranges. This is mainly because, in such a high-temperature shock layer where the temperature of interest is in the order of 5000 K to 10,000 K, as shown in Fig.~\ref{Bp_RadiationCoupling_Contour}, the peak radiation density is located close to the VUV region, causing the most perceptive change of the radiation intensity as a response to the temperature changes due to the radiation coupling. It is worth clarifying that the LOS-1, LOS-2, and LOS-3 correspondingly present 14.9\%, 16.5\%, and 16.6\% reductions of the radiative heat flux values owing to the radiation coupling if the overall wavelength range (\emph{i.e.}, from VUV to IR) is considered. This proposes evidence of the importance of considering the coupling effect in the multi-dimensional configuration for the accurate characterization of the aerothermal environment, unlike the conventional existing studies that mostly focus on stagnation line analyses.

% Figure-11 in MS Word

\begin{figure}[h!]
    \centering
    \includegraphics[width=0.55\textwidth]{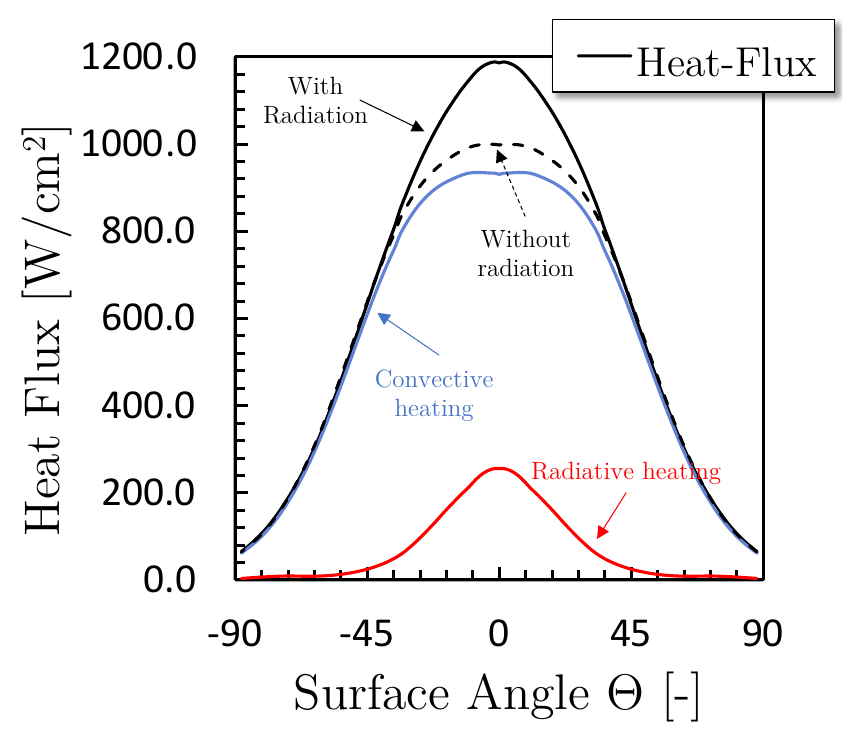}
    \caption{Comparison of the surface heat flux in case of the B’ equilibrium ablation model with and without radiative transfers.}
    \label{Bp_RadiationCoupling_Wall}
\end{figure}

When observing the vehicle surface, the influence of a multi-physics coupling treatment on hypersonic entry characterization has been highlighted by the various existing coupling strategies previously achieved in the literature. As discussed by Johnston~\cite{Johnston2014}, considering radiative transfer decreases the incoming convective heat flux over the surface. However, summing the radiative and convective heat fluxes brings back to the total amount of energy transferred at the wall above the CFD-Material coupled case (see Fig. \ref{Park5_Bp_Contour}) as shown in Fig. \ref{Bp_RadiationCoupling_Wall}. Comparing the surface heat flux obtained with and without radiative transfers, a difference of up to 26\% at the stagnation point is observed.\\ 

% Figure-12 in MS Word

\begin{figure}[h!]\label{fig13}
    \centering
    \includegraphics[width=1.0\textwidth]{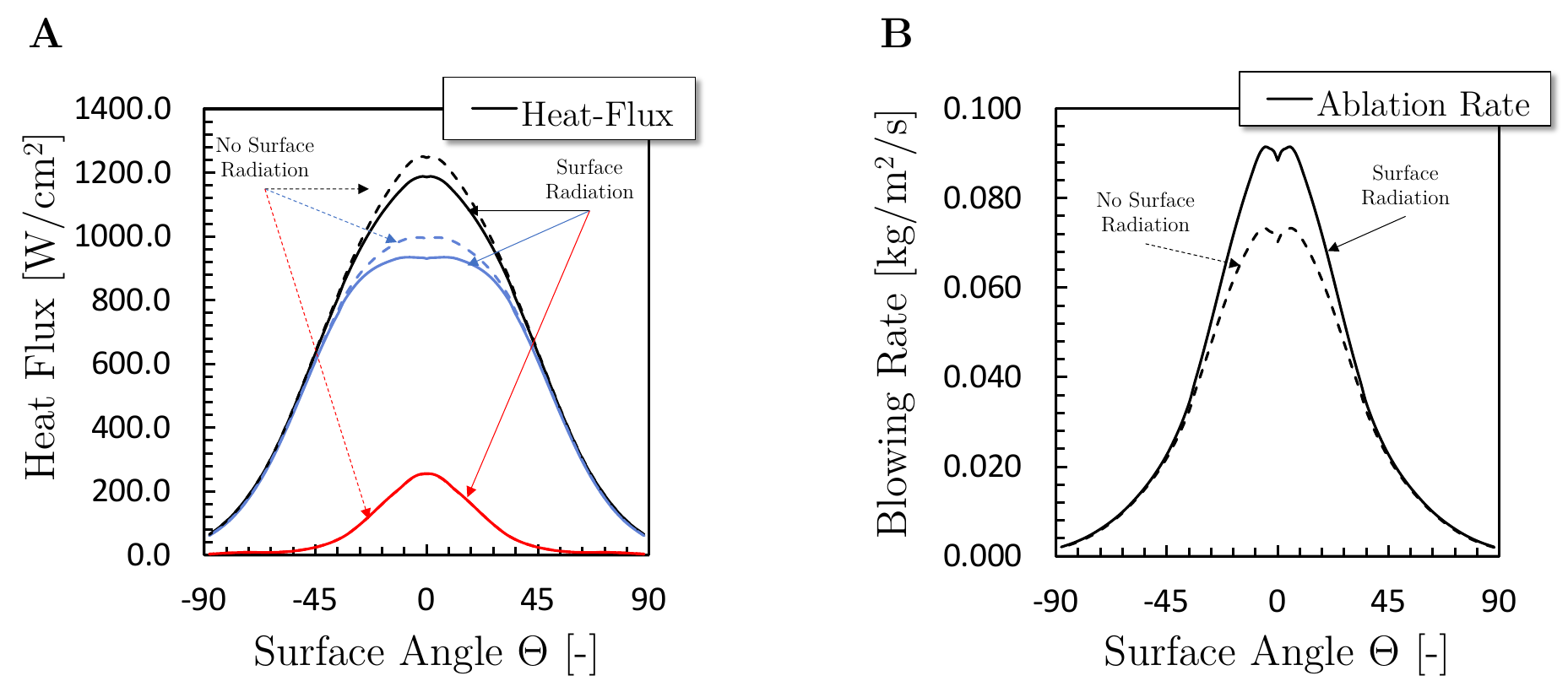}
    \caption{Influence of the radiative surface coupling on the surface state. (A) Surface heat flux where the black, blue, and red lines correspond to total, convective, and radiative heat fluxes. (B) Blowing rate.}
    \label{Bp_SurfaceRadiationCoupling}
\end{figure}

As described in the surface equilibrium equation, the blowing rate is directly scaled by the total heat flux (\emph{i.e.}, summation of radiative and convective contributions). Accounting for radiative heat flux over the surface is, therefore, raising the blowing rate on the vehicle surface and leads to a non-negligible reduction of convective heat flux when compared to cases where the material solver is only coupled to the hypersonic flows \cite{Jo2023b}, and radiative heating is evaluated \emph{a posteriori} in an uncoupled manner. The differences observed between the cases are highlighted in Fig.~\ref{Bp_SurfaceRadiationCoupling}.

% Figure-13 in MS Word

\begin{figure}[h!]\label{fig14}
    \centering
    \includegraphics[width=1.0\textwidth]{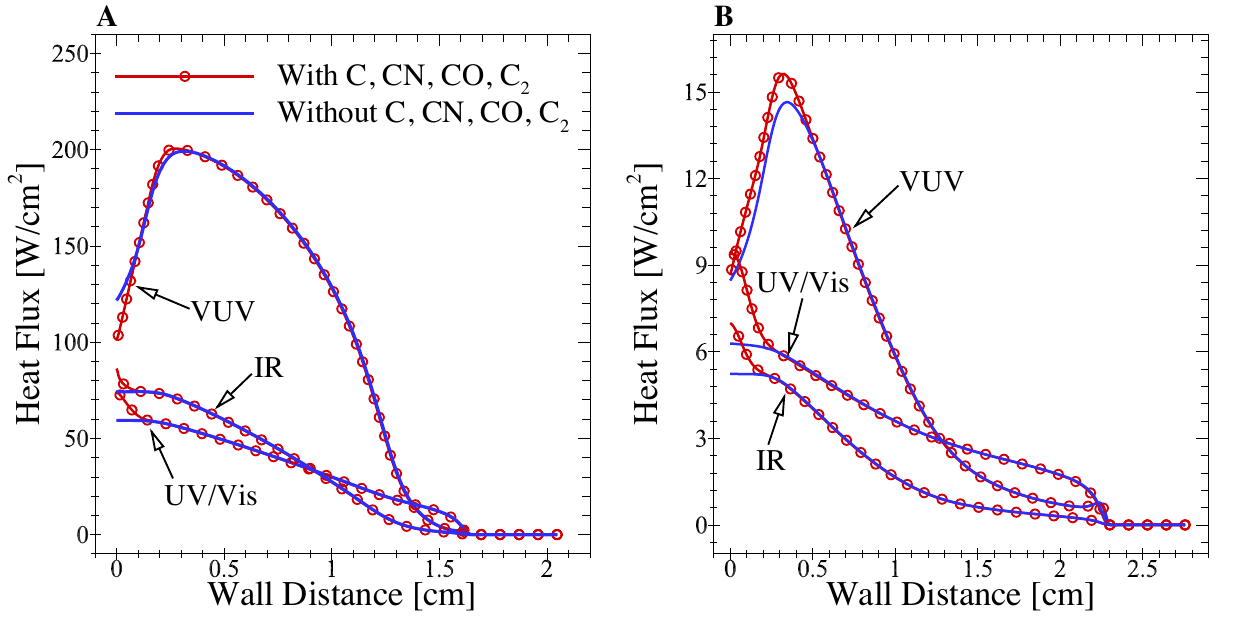}
    \caption{Influence of the ablation species on the radiative heat flux profiles. (A) LOS-1. (B) LOS-2.}
    \label{Qrad_AblationEffect}
\end{figure}

Accounting for the radiative heat flux over the material in a coupled manner seems to mainly influence the blowing rate (c.f. Fig. \ref{Bp_SurfaceRadiationCoupling}-B) and participates in the convective heat flux reduction as depicted in Fig. \ref{Bp_SurfaceRadiationCoupling}-A. Exclusion of radiative transfers around the vehicle surface coupling leads to over-prediction of the total heat flux up to 5\% at the stagnation point and an under-prediction of the ablation rate of the material of 20\%. A slight variation of the radiative heat flux is observable in Fig. \ref{Bp_SurfaceRadiationCoupling}-A, demonstrating a relatively weak coupling between the blowing rate and the radiative heating for the predicted surface temperature. To characterize the contributions of the ablation species to the radiative heat flux, the previous case is repeated while discarding radiative contributions of C, CO, CN, and C$_2$. Figure \ref{Qrad_AblationEffect} shows the changes in radiative heat flux due to ablation species along LOS-1 and LOS-2. As shown in Fig. \ref{Qrad_AblationEffect}-A, The ablation species absorb the radiation in the VUV wavelength range, whereas they emit in the UV/Vis and IR regions. About 20\% of the reduction in radiative heat flux is observed in the VUV due to the severe absorption, which is mainly attributed to the CO 4th positive system.
In contrast, the increase in UV/Vis and IR is due to CN, C, and C$_2$. In Fig.~\ref{Qrad_AblationEffect}(B), it is important to note that the ablation species increase the radiative heat flux even in the VUV range, unlike the trend along the LOS-1 and in the previous study~\cite{Park2001} in which the stagnation line analysis was carried out. This may imply that the multi-dimensional effect is critical for an accurate characterization of the aerothermal heating of hypersonic vehicles. This is also strong evidence demonstrating the necessity of the multi-solver coupled approach in high-fidelity, as developed in the present study, to predict the trend along the off-stagnation regions.

% Figure-14 in MS Word

\section{Conclusions}\label{Sec:Conclusion}
In this study, a multi-solver approach has been used to efficiently analyze the multi-physics coupled aspect of the Earth's atmospheric entry in a strongly coupled manner. The radiative source, computed from local fluid properties, is injected directly into the energy conservation equations of the hypersonic flow solver and on the vehicle surface energy balance. An equilibrium ablation model has been investigated in depth to study the influence of ablation species on radiative heating, as well as the coupled solution sensibility regarding the main parameter of the ablation modeling, which is the Stanton number in the considered case. Comparison with the representative finite-rate model obtained from the literature has shown that the char production rate is affected by almost a factor of three regarding the equilibrium model predictions. This observation has permitted to address the solution sensibility owing the finite-rate nature of the ablation model for the observed flight point through a scaling coefficient in the equilibrium model, simplifying the analysis of fully coupled framework for the rest of this study while offering better numerical stability.

By performing these multi-solver simulations, the importance of multi-physics considerations for high-speed atmospheric entry has been demonstrated. The inclusion of the radiative heat flux along the ablative material surface into the surface energy balance increases the blowing rate up to 20\%. The strong coupling among the three solvers is critical for the accurate characterization of aerothermal heating and the ablation rate. The ablative products strongly absorb the radiative heat flux in the VUV range along the stagnation line while emitting in the longer wavelength range. The ablative species emit radiative energy in the off-stagnation line, increasing the surface heat flux. This demonstrates the necessity of developing the multi-solver coupled framework that the present work proposes. In future studies, the material response is aimed to be more realistic for applications accounting for mass and energy transport inside a porous pyrolytic material. This will necessitate to account for additional species, both from a chemical-kinetics and radiative properties point of view. Moreover, additional finite-rate ablation models could be investigated. These models usually rely on additional empirical parameters whose sensibilities also need to be evaluated to improve the predictive capabilities of the current numerical framework. 

\section*{Acknowledgments}
The authors gratefully acknowledge the support of NASA through the NASA Space Technology Research Institute (NSTRI) ACCESS, under Grant No. 80NSSC21K1117.

\section{Appendix}\label{Sec:Appendix}

\begin{landscape}
\begin{table}[h!]
    \centering
    \begin{tabular}{|c|c|c|c|}
        \hline
        \textbf{Mechanism} & \textbf{Type} & \textbf{Corresponding Ablation Products Expression} & \textbf{Parameters} \\
        \hline
        $\text{O} + \text{C(s)} \rightarrow \text{CO} + (\text{s})$ & \makecell{\\[12pt] Eley-Rideal \\[12pt]} & $s_{\text{CO}}^s = \rho^g y_{\text{O}}^g \gamma_1 \sqrt{\frac{RT}{2\pi M_{\text{O}}}} \frac{M_{\text{CO}}}{M_{\text{O}}}$ & $\gamma_1 = 0.63 e^{\frac{-1160}{T}}$ \\
        \hline
        $\text{O}_2 + 2\text{C(s)} \rightarrow 2\text{CO} + (\text{s})$ & \makecell{\\[12pt] Eley-Rideal \\[12pt]} & $s_{\text{CO}}^s = 2\rho^g y_{\text{O}_2}^g \gamma_2 \sqrt{\frac{RT}{2\pi M_{\text{O}_2}}} \frac{M_{\text{CO}}}{M_{\text{O}_2}}$ & $\gamma_2 = \frac{0.00143 + 0.01 e^{\frac{-1450}{T}}}{1 + 0.0002 e^{\frac{13000}{T}}}$ \\
        \hline
        $\text{N} + \text{C(s)} \rightarrow \text{CN} + (\text{s})$ & \makecell{\\[12pt] Eley-Rideal \\[12pt]} & $s_{\text{CN}}^s = \rho^g y_{\text{N}}^g \gamma_3 \sqrt{\frac{RT}{2\pi M_{\text{N}}}} \frac{M_{\text{CN}}}{M_{\text{N}}}$ & $\gamma_3 = 0.001$ \\
        \hline
        $\text{N} + \text{N(s)} \rightarrow \text{N}_2$ & \makecell{\\[12pt] Eley-Rideal \\[12pt]} & $s_{\text{N}_2}^s = \rho^g y_{\text{N}}^g \gamma_4 \sqrt{\frac{RT}{2\pi M_{\text{N}}}}$ & $\gamma_4 = 0.05$ \\
        \hline
        $3\text{C(s)} \rightarrow \text{C}_3$ & Sublimation & $s_{\text{C}_3}^s = \sqrt{\frac{RT}{2\pi M_{\text{C}_3}}} \alpha_{\text{C}_3} \left( A_{\text{C}_3} T^{N_{\text{C}_3}} e^{\frac{-E_{\text{C}_3}}{T}} - \rho^g y_{\text{C}_3}^g \right)$ & \makecell{\\[12pt] $\alpha_{\text{C}_3} = 0.03$ \\ $A_{\text{C}_3} = 4.3194 \times 10^{22}$ \\ $N_{\text{C}_3} = -3.459$ \\ $E_{\text{C}_3} = 103339$ \\[12pt]}\\
        \hline
    \end{tabular}
    \caption{Park-5 ablation model parameters used in this study.}
    \label{table:Park-5}
\end{table}
\end{landscape}
\newpage
%%%%%%%%%%%%%%%%%%%%%%%%%%%%%%%%%%%%%%%%%%%%%%%%%%%%%%%%%%%
% Figure bank 
%%%%%%%%%%%%%%%%%%%%%%%%%%%%%%%%%%%%%%%%%%%%%%%%%%%%%%%%%%%

%%%%%%%%%%%%%%%%%%%%%%%%%%%%%%%%%%%%%%%%%%%%%%%%%%%%%%%%%%%

%% The Appendices part is started with the command \appendix;
%% appendix sections are then done as normal sections
%% \appendix

%% \section{}
%% \label{}

%% If you have bibdatabase file and want bibtex to generate the
%% bibitems, please use
%%
%%  \bibliographystyle{elsarticle-num} 
%%  \bibliography{<your bibdatabase>}

%% else use the following coding to input the bibitems directly in the
%% TeX file.
%\bibliographystyle{plainnat}  %% Use an author-year citation style
\bibliographystyle{unsrtnat}

\bibliography{references}  %% Name of your .bib file without the .bib extension

\begin{thebibliography}{35}
\providecommand{\natexlab}[1]{#1}
\providecommand{\url}[1]{\texttt{#1}}
\expandafter\ifx\csname urlstyle\endcsname\relax
  \providecommand{\doi}[1]{doi: #1}\else
  \providecommand{\doi}{doi: \begingroup \urlstyle{rm}\Url}\fi

\bibitem[Barbante and Magin(2004)]{Barbante2004}
P.~F. Barbante and T.~E. Magin.
\newblock Fundamentals of hypersonic flight - properties of high temperature gases.
\newblock In \emph{Critical Technologies for Hypersonic Vehicle Development}, pages 1--50. NATO-STO, Rhode-St-Genèse, 2004.

\bibitem[Gnoffo(1999)]{Gnoffo1999}
Peter~A. Gnoffo.
\newblock Planetary-entry gas dynamics.
\newblock \emph{Annual Review of Fluid Mechanics}, 31:\penalty0 459--494, 1999.
\newblock \doi{10.1146/annurev.fluid.31.1.459}.

\bibitem[Josyula and Vedula(2015)]{Josyula2015}
E.~Josyula and P.~Vedula.
\newblock Fundamental fluid transport equations for hypersonic nonequilibrium flows.
\newblock In \emph{Hypersonic Nonequilibrium Flows: Fundamentals and Recent Advances}. American Institute of Aeronautics and Astronautics, VA, USA, 2015.

\bibitem[Park(1990)]{Park1990}
Chul Park.
\newblock \emph{Nonequilibrium hypersonic aerothermodynamics}.
\newblock Wiley, New York, 1990.
\newblock ISBN 9780471510932.

\bibitem[Johnston et~al.(2008)Johnston, Hollis, and Sutton]{Johnston2008}
C.~O. Johnston, B.~R. Hollis, and K.~Sutton.
\newblock Spectrum modeling for air shock-layer radiation at lunar-return conditions.
\newblock \emph{Journal of Spacecraft and Rockets}, 45\penalty0 (5):\penalty0 865--878, 2008.

\bibitem[Marschall et~al.(2015)Marschall, MacLean, Norman, and Schwartzentruber]{Marschall2015}
J.~Marschall, M.~MacLean, P.~E. Norman, and T.~E. Schwartzentruber.
\newblock Surface chemistry in non-equilibrium flows.
\newblock In \emph{Hypersonic Nonequilibrium Flows: Fundamentals and Recent Advances}, pages 239--327. American Institute of Aeronautics and Astronautics, 2015.

\bibitem[Uyanna and Najafi(2020)]{Uyanna2020}
O.~Uyanna and H.~Najafi.
\newblock Thermal protection systems for space vehicles: A review on technology development, current challenges and future prospects.
\newblock \emph{Acta Astronautica}, 176:\penalty0 341--356, 2020.

\bibitem[Meurisse et~al.(2018)Meurisse, Lachaud, Panerai, Tang, and Mansour]{Meurisse2018}
J.~B. Meurisse, J.~Lachaud, F.~Panerai, C.~Tang, and N.~N. Mansour.
\newblock Multidimensional material response simulations of a full-scale tiled ablative heatshield.
\newblock \emph{Aerospace Science and Technology}, 76:\penalty0 497--511, 2018.

\bibitem[Johnston et~al.(2012)Johnston, Brandis, and Sutton]{Johnston2012}
C.~O. Johnston, A.~M. Brandis, and K.~Sutton.
\newblock Shock layer radiation modeling and uncertainty.
\newblock In \emph{43rd AIAA Thermophysics Conference}, pages 1--43, New Orleans, 2012. American Institute of Aeronautics and Astronautics.

\bibitem[Park(2001)]{Park2001}
C.~Park.
\newblock Stagnation-point radiation for {Apollo} 4 - a review and current status.
\newblock In \emph{35th AIAA Thermophysics Conference}, pages 1--17, Anaheim, 2001. American Institute of Aeronautics and Astronautics.

\bibitem[Moss(1976)]{Moss1976}
J.~N. Moss.
\newblock Radiative viscous-shock-layer solutions with coupled ablation injection.
\newblock \emph{AIAA Journal}, 14\penalty0 (9):\penalty0 1311--1317, 1976.

\bibitem[Johnston and Gnoffo(2009)]{Johnston2009}
C.~O. Johnston and P.~A. Gnoffo.
\newblock Influence of ablation on radiative heating for {Earth} entry.
\newblock \emph{Journal of Spacecraft and Rockets}, 46\penalty0 (3):\penalty0 481--491, 2009.

\bibitem[Olynick et~al.(1999)Olynick, Chen, and Tauber]{Olynick1999}
D.~Olynick, Y.-K. Chen, and M.~E. Tauber.
\newblock Aerothermodynamics of the {Stardust} sample return capsule.
\newblock \emph{Journal of Spacecraft and Rockets}, 36\penalty0 (3):\penalty0 442--462, 1999.

\bibitem[Johnston(2014)]{Johnston2014}
C.~O. Johnston.
\newblock Study of aerothermodynamic modeling issues relevant to high-speed sample return vehicles.
\newblock In \emph{NATO-STO}. Rhode-St-Genèse, 2014.

\bibitem[Doihara and Nishida(2004)]{Doihara2004}
R.~Doihara and M.~Nishida.
\newblock Ablation studies for a super-orbital reentry capsule using a three-temperature model.
\newblock \emph{Japan Society for Aeronautical and Space Sciences}, 47\penalty0 (157):\penalty0 161--166, 2004.

\bibitem[{National Aeronautics and Space Administration}(2015)]{NASA2015}
{National Aeronautics and Space Administration}.
\newblock {NASA} technology roadmaps {TA} 14: Thermal management systems.
\newblock Technical report, 2015.

\bibitem[Candler(2012)]{Candler2012}
G.~Candler.
\newblock Nonequilibrium processes in hypervelocity flows: An analysis of carbon ablation models.
\newblock In \emph{50th AIAA Aerospace Sciences Meeting}, pages 1--19, Nashville, 2012. American Institute of Aeronautics and Astronautics.

\bibitem[Padovan et~al.(2024)Padovan, Vollmer, Panerai, Panesi, Stephani, and Bodony]{padovan2024extended}
A.~Padovan, B.~Vollmer, F.~Panerai, M.~Panesi, K.~A. Stephani, and D.~J Bodony.
\newblock An extended {B}' formulation for ablating-surface boundary conditions.
\newblock \emph{International Journal of Heat and Mass Transfer}, 218:\penalty0 124770, 2024.

\bibitem[Chiodi et~al.(2022)Chiodi, Stephani, Panesi, and Bodony]{chiodi2022chyps}
R.~M Chiodi, K.~A Stephani, M.~Panesi, and D.~J Bodony.
\newblock {CHyPS}: A high-order material response solver for ablative thermal protection systems.
\newblock In \emph{AIAA SCITECH 2022 Forum}, page 1501, 2022.

\bibitem[Gnoffo et~al.(2010)Gnoffo, Johnston, and Thompson]{Gnoffo2010}
P.~A. Gnoffo, C.~O. Johnston, and R.~A. Thompson.
\newblock Implementation of radiation, ablation, and free energy minimization in hypersonic simulations.
\newblock \emph{Journal of Spacecraft and Rockets}, 47\penalty0 (2):\penalty0 251--257, 2010.

\bibitem[Jo et~al.(2023{\natexlab{a}})Jo, Maout, Munafò, and Panesi]{Jo2023b}
S.~M. Jo, V.~Le Maout, A.~Munafò, and M.~Panesi.
\newblock A multi-solver approach for studying ablation and radiation interactions in hypersonic flows.
\newblock In \emph{AIAA Aviation 2023 Forum}, pages 1--15, San Diego, 2023{\natexlab{a}}. American Institute of Aeronautics and Astronautics.

\bibitem[Munafò et~al.(2020)Munafò, Alberti, Pantano, Freund, and Panesi]{Munafo2020}
A.~Munafò, A.~Alberti, C.~Pantano, J.~B. Freund, and M.~Panesi.
\newblock A computational model for nanosecond pulse laser-plasma interactions.
\newblock \emph{Journal of Computational Physics}, 406:\penalty0 109190, 2020.

\bibitem[Jo et~al.(2023{\natexlab{b}})Jo, Kumar, Maout, Munafò, and Panesi]{Jo2023a}
S.~M. Jo, S.~Kumar, V.~Le Maout, A.~Munafò, and M.~Panesi.
\newblock Multi-fidelity modeling framework for radiative transfer in hypersonic atmospheric entry.
\newblock In \emph{AIAA Scitech 2023 Forum}, pages 1--13, Washington DC, 2023{\natexlab{b}}. American Institute of Aeronautics and Astronautics.

\bibitem[Chourdakis et~al.(2022)Chourdakis, Davis, Rodenberg, Schulte, Simonis, Uekermann, et~al.]{Chourdakis2022}
G.~Chourdakis, K.~Davis, B.~Rodenberg, M.~Schulte, F.~Simonis, B.~Uekermann, et~al.
\newblock precice v2: A sustainable and user-friendly coupling library.
\newblock \emph{Open Research Europe}, 2\penalty0 (51), 2022.

\bibitem[Johnston et~al.(2018)Johnston, Sahai, and Panesi]{Johnston2018}
C.~O. Johnston, A.~Sahai, and M.~Panesi.
\newblock Extension of multiband opacity-binning to molecular, non-boltzmann shock layer radiation.
\newblock \emph{Journal of Thermophysics and Heat Transfer}, 32\penalty0 (3):\penalty0 816--821, 2018.

\bibitem[Scoggins et~al.(2013)Scoggins, Magin, Wray, and Mansour]{Scoggins2013}
J.~B. Scoggins, T.~Magin, A.~A. Wray, and N.~Mansour.
\newblock Multi-group reductions of lte air plasma radiative transfer in cylindrical geometries.
\newblock In \emph{44th AIAA Thermophysics Conference}, pages 1--18, San Diego, 2013. American Institute of Aeronautics and Astronautics.

\bibitem[Munafò and Panesi(2023)]{Munafo2023}
A.~Munafò and M.~Panesi.
\newblock Plato: A high-fidelity tool for multi-component plasmas.
\newblock In \emph{AIAA Aviation 2023 Forum}, San Diego, 2023. American Institute of Aeronautics and Astronautics.

\bibitem[Martin et~al.(2017)Martin, Zhang, and Tagavi]{Martin2017}
A.~Martin, H.~Zhang, and K.~A. Tagavi.
\newblock An introduction to the derivation of surface balance equations without the excruciating pain.
\newblock \emph{International Journal of Heat and Mass Transfer}, 115:\penalty0 992--999, 2017.

\bibitem[Thompson and Gnoffo(2012)]{Thompson2012}
R.~Thompson and P.~Gnoffo.
\newblock Implementation of a blowing boundary condition in the laura code.
\newblock In \emph{46th AIAA Aerospace Sciences Meeting and Exhibit}, pages 1--11, Reno, 2012. American Institute of Aeronautics and Astronautics.

\bibitem[Martin and Boyd(2011)]{Martin2011}
A.~Martin and I.~D. Boyd.
\newblock {CFD} implementation of a novel carbon-phenolic-in-air chemistry model for atmospheric re-entry.
\newblock In \emph{49th AIAA Aerospace Sciences Meeting}, pages 1--19, Orlando, 2011. American Institute of Aeronautics and Astronautics.

\bibitem[Maout et~al.(2023)Maout, Munafò, and Panesi]{LeMaout2023}
V.~Le Maout, A.~Munafò, and M.~Panesi.
\newblock Effects of non equilibrium surface boundary conditions for material response in atmospheric reentry simulations.
\newblock In \emph{AIAA Scitech 2023 Forum}, pages 1--12, Washington DC, 2023. American Institute of Aeronautics and Astronautics.

\bibitem[Scoggins et~al.(2017)Scoggins, Rabinovitch, Barros-Fernandez, Martin, Lachaud, Jaffe, et~al.]{Scoggins2017}
J.~B. Scoggins, J.~Rabinovitch, B.~Barros-Fernandez, A.~Martin, J.~Lachaud, R.~L. Jaffe, et~al.
\newblock Thermodynamic properties of carbon–phenolic gas mixtures.
\newblock \emph{Aerospace Science and Technology}, 66:\penalty0 177--192, 2017.

\bibitem[Schlichting and Gersten(2017)]{Schlichting2017}
H.~Schlichting and K.~Gersten.
\newblock \emph{Boundary-Layer Theory}.
\newblock Springer, Berlin, 2017.

\bibitem[Martin and Boyd(2014)]{Martin2014}
A.~Martin and I.~D. Boyd.
\newblock Strongly coupled computation of material response and nonequilibrium flow for hypersonic ablation.
\newblock \emph{Journal of Spacecraft and Rockets}, 52\penalty0 (1):\penalty0 1--16, 2014.

\bibitem[Jo et~al.(2020)Jo, Kwon, and Kim]{Jo2020}
S.~M. Jo, O.~J. Kwon, and J.~G. Kim.
\newblock Stagnation-point heating of {Fire II} with a non-{B}oltzmann radiation model.
\newblock \emph{International Journal of Heat and Mass Transfer}, 153:\penalty0 1--15, 2020.

\end{thebibliography}

%\begin{thebibliography}{00}

%% \bibitem{label}
%% Text of bibliographic item

%\bibitem{}

%\end{thebibliography}
\end{document}